\documentclass[sigconf, screen,table]{acmart}

\acmConference[]{}
\acmPrice{}
\acmISBN{}

\renewcommand\footnotetextcopyrightpermission[1]{}
\setcopyright{none}

\newcommand{\NAME}{etylizer}
\usepackage[english]{babel}
\usepackage{listings}
\usepackage{url}
\usepackage{csquotes}
\usepackage{amsmath}
\usepackage{svg}
\usepackage{marginnote}
\usepackage{stmaryrd}
\usepackage{placeins}
\usepackage{tabularx}
\usepackage{mathpartir}
\usepackage{ifthen}
\usepackage{balance}
\usepackage{enumerate}
\usepackage[colorinlistoftodos]{todonotes}
\setuptodonotes{inline}

\definecolor{GrayBgColor}{rgb}{0.9, 0.9, 0.9}
\definecolor{GrayFgColor}{rgb}{0.4, 0.4, 0.4}
\definecolor{StringColor}{rgb}{0.0, 0.38039, 0.141176}
\usepackage{graphicx}
\usepackage{cleveref} % automically spells out "Figure" or "Section" when using \label
\crefname{figure}{Fig.}{Figs.}
\crefname{section}{Sec.}{Secs.}
\crefname{subsection}{Sec.}{Secs}
\crefname{subsubsection}{Sec.}{Secs.}
\crefname{appendix}{App.}{Apps.}
\usepackage{pgfplots, pgfplotstable}
\usetikzlibrary{patterns}

\pgfplotsset{width=7cm,compat=1.9}

\renewcommand\emph[1]{\textit{#1}}

\newcommand\MinErl{\ensuremath{\lambda_{\textsf{erl}}}}

\newcommand*{\medcup}{\mathbin{\scalebox{0.8}{\ensuremath{\bigcup}}}}%
\newcommand*{\medwedge}{\mathbin{\scalebox{0.8}{\ensuremath{\bigwedge}}}}%
\newcommand*{\medvee}{\mathbin{\scalebox{0.8}{\ensuremath{\bigvee}}}}%
\newcommand{\doublewedge}{%
  \mathbin{
    \mathchoice{\wedge\mkern-15mu\wedge}
               {\wedge\mkern-15mu\wedge}
               {\wedge\mkern-12.5mu\wedge}
               {\wedge\mkern-11mu\wedge}
    }
}

\newcommand{\figureboxSingle}[1]
        {\fbox{\begin{minipage}{\columnwidth} #1 \end{minipage}}}
\newcommand{\boxfigSingle}[3]
           {\begin{figure}\figureboxSingle{#3}\vspace{-2ex}\caption{\label{#1}#2}\end{figure}}

\newcommand\Multi[2][!*NEVER USED ARGUMENT*!]{%
  \ifthenelse{\equal{#1}{!*NEVER USED ARGUMENT*!}}{#2_{i \in I}}{#2_{#1}}%
}
\newcommand{\kw}[1]{\ensuremath{\mathtt{#1}}}
\newcommand\Name[1]{\ensuremath{\mathtt{#1}}}

\newcommand\Rule[1]{\TirName{#1}}
\def\RuleForm#1{{\setlength{\fboxrule}{0.5pt}\fbox{\normalsize \ensuremath{#1}}}}
\newcommand\RuleSection[2]{%
  \begin{tabularx}{\textwidth}{Xr}%
    #1\hfill & {\small \it #2}%
  \end{tabularx}%
}

\makeatother
\lstdefinelanguage{erlang}{%
   morekeywords={after,and,case,catch,div,end,exit,export,if,import,module,of,or,%
     receive,rem,-spec,throw,when%
   },
   otherkeywords={-spec,-type},%
   xleftmargin=0.1cm,%
   numbersep=3pt,
   sensitive,%
   basicstyle=\lst@ifdisplaystyle\scriptsize\footnotesize\fi\ttfamily,
   keywordstyle=\color{black}\bf\ttfamily,
   numberstyle=\tiny\color{GrayFgColor}\sffamily\raisebox{0.6pt},
   morecomment=[n]{\{-}{-\}},%
   morestring=[b]",%
  }[keywords,comments,strings]%
\makeatletter

\newcommand\IsSubty{\leq}
\newcommand\IsSubtyBase{\leq_{\mathsf{b}}}
\newcommand\ValMatches{\mathbin{\#}}

\newcommand\MoreGeneral\sqsubseteq
\newcommand\Inter{\wedge}
\newcommand\Union{\vee}
\newcommand\InterBig{\medwedge}
\newcommand\UnionBig{\medvee}
\newcommand\WithoutTy\setminus

\newcommand\Pairty[2]{#1 \times #2}

\newcommand\ErlTop{\top}
\newcommand\ErlBot{\bot}
\newcommand\Neg{\neg}
\newcommand\Atom{\Name{atom}}
\newcommand\Int{\Name{int}}
\newcommand\Float{\Name{float}}

\newcommand\AnyPair{\Name{pair}}
\newcommand\AnyFun{\Name{fun}}

\newcommand\Polyty{\sigma}
\newcommand\Monoty{t}
\newcommand\Tyvar{\alpha}
\newcommand\TyvarAux{\beta}
\newcommand\TyvarSet{A}
\newcommand\TyScm[1]{\forall #1\,.\,}
\newcommand\Constty{\mathbb{K}}
\newcommand\Basety{b}

\newcommand\LetrecSym{\kw{letrec}}
\newcommand\Letrec[2]{\LetrecSym~#1~\kw{in}~#2}
\newcommand\Case[1]{\kw{case}~#1~\kw{of}~}

\newcommand\DefSym{\mathit{def}}

\newcommand\Wildcard{\_}
\newcommand\EqSynSym{\mathbin{\texttt{=}}}
\newcommand\Def[3]{#1 \mathop{\texttt{:}} #2 \EqSynSym #3}
\newcommand\DefNt[2]{#1 \EqSynSym #2}
\newcommand\Abs[1]{\lambda #1 \texttt{.}}
\newcommand\App[1]{#1\,}
\newcommand\Pair[2]{\texttt{(}#1\texttt{,} #2\texttt{)}}
\newcommand\When{\mathrel{\kw{when}}}
\newcommand\Cls{\mathit{cls}}
\newcommand\Pg{\mathit{pg}}
\newcommand\WithGuard[2]{#1 \When #2}
\newcommand\PatCls[2]{#1 \to #2}
\newcommand\Is[2]{\Name{is}_{#1}(#2)}
\newcommand\Const{\kappa}
\newcommand\GuardOracle{\kw{oracle}}
\newcommand\GuardAndSym{\mathop{\kw{and}}}
\newcommand\GuardAnd[2]{#1 \GuardAndSym #2}
\newcommand\GuardTrue{\kw{true}}

\newcommand\EmptyEnv{\{\}}
\newcommand\Dom[1]{\mathsf{dom}(#1)}
\newcommand\Forall[1]{(\forall #1)}
\newcommand\Free[1]{\mathsf{fv}(#1)}
\newcommand\Bound[1]{\mathsf{bv}(#1)}
\newcommand\FreeTyVars[1]{\mathsf{ftv}(#1)}
\newcommand\EnvsSym{\mathsf{envs}}
\newcommand\Envs[2]{\EnvsSym_{#1}(#2)}

\newcommand\Angle[1]{\langle#1\rangle}
\newcommand\MetaPair[2]{\Angle{#1; #2}}
\newcommand\Tysubst{\theta}
\newcommand\Valsubst{\eta}
\newcommand\GenSym{\mathsf{gen}}
\newcommand\Gen[2]{\GenSym_{#1}(#2)}
\newcommand\EquivSym{\mathsf{equiv}}
\newcommand\Equiv[1]{\EquivSym(#1)}

\newcommand\ReduceSym{\leadsto}
\newcommand\Reduce[3]{#1 \vdash #2 \ReduceSym #3}
\newcommand\PatSubst[2]{#1/#2}
\newcommand\FunEnv{\Valsubst}
\newcommand\PatSubstFail{\Omega}
\newcommand\EvalCtx{\mathcal E}
\newcommand\Hole{\scalebox{0.8}{\ensuremath{\square}}}

\newcommand\ValEq{\equiv}
\newcommand\SubstUnion\oplus
\newcommand\Plug[2]{#1[#2]}

\newcommand\EvalGuard[1]{\mathcal{G}(#1)}
\newcommand\True{\mathit{true}}
\newcommand\False{\mathit{false}}

\newcommand\Ok{\mathsf{ok}}
\newcommand\ExpTy[3]{#1 \Turns #2 : #3}

\newcommand\DefOk[2]{#1 \Turns #2 ~\Ok}

\newcommand\Turns[1][!*NEVER USED ARGUMENT*!]{%
  \ifthenelse{\equal{#1}{!*NEVER USED ARGUMENT*!}}{\vdash}{\vdash_{\mathsf{#1}}}%
}
\newcommand\Venv{\Gamma}
\newcommand\PgUpperTy[2]{\lceil #1 \rceil_{#2}}
\newcommand\PgLowerTy[2]{\lfloor #1 \rfloor_{#2}}
\newcommand\PatTy[2]{\lbag #1 \rbag_{#2}}

\newcommand\TyOfConstSym{\mathsf{ty}}
\newcommand\TyOfConst[1]{\TyOfConstSym(#1)}
\newcommand\PatEnv[2]{#1\sslash#2}
\newcommand\ProjL[1]{\pi_1(#1)}
\newcommand\ProjR[1]{\pi_2(#1)}
\newcommand\InterEnv{\doublewedge}
\newcommand\PatFromExp[1]{\llparenthesis\, #1\, \rrparenthesis}
\newcommand\Inst[1]{\InstSym(#1)}
\newcommand\InstSym{\mathsf{inst}}
\newcommand\Env[1]{\mathsf{env}(#1)}

\newcommand\Constr{c}
\newcommand\ConstrSet{C}
\newcommand\SiConstr{d}
\newcommand\SiConstrSet{D}
\newcommand\IsSubtyConstr{\mathbin{\dot{\IsSubty}}}
\newcommand\SubtyConstr[2]{#1 \IsSubtyConstr #2}
\newcommand\DefConstr[2]{\kw{def}~#1~\kw{in}~#2}
\newcommand\LetConstr[3]{\kw{letrec}\,\MetaPair{#1}{#2}~\kw{in}~#3}
\newcommand\CaseConstr[2]{\kw{case}~#1~\kw{of}~#2}
\newcommand\InConstr[3]{#1~\kw{in}~#2~\kw{when}~#3}
\newcommand\ConstrGen[3]{#1 : #2 \Rightarrow #3}
\newcommand\DefConstrGen[3]{#1 \Rightarrow \MetaPair{#2}{#3}}
\newcommand\ConstrRew[3]{#1 \Turns #2 \leadsto #3}
\newcommand\PatTyEnvConstr[4]{#1 \sslash #2 \Rightarrow \MetaPair{#3}{#4}}

\newcommand\TallySym{\mathsf{tally}}
\newcommand\Tally[2][!*NEVER USED ARGUMENT*!]{%
  \ifthenelse{\equal{#1}{!*NEVER USED ARGUMENT*!}}{\TallySym(#2)}{\TallySym_{#1}(#2)}%
}

\newcommand\Model{\ensuremath{\mathcal M}}
\newcommand\Filtermap{\mathit{filtermap}}
\newcommand\AtomSingle[1]{\ensuremath{\mathsf{#1}}}
\newcommand\AtomTrue{\AtomSingle{true}}
\newcommand\AtomFalse{\AtomSingle{false}}
\newcommand\AtomNil{\AtomSingle{nil}}
\newcommand\ListTy[1]{\mathit{list}\,#1}

\begin{document}

%% List of approved packages: https://authors.acm.org/proceedings/production-information/accepted-latex-packages

\title{Set-theoretic Types for Erlang}

\author{Albert Schimpf}
\email{schimpf@cs.uni-kl.de}
%\orcid{}
\affiliation{%
   \institution{University of Kaiserslautern-Landau}	
  \country{Germany}
}

\author{Stefan Wehr}
\email{stefan.wehr@hs-offenburg.de}
\affiliation{%
  \institution{Offenburg Univ.\ of Applied Sciences}
  \country{Germany}
}

\author{Annette Bieniusa}
\email{bieniusa@cs.uni-kl.de}
\affiliation{%
  \institution{University of Kaiserslautern-Landau}	
  \country{Germany}
}


\begin{abstract}
  Erlang is a functional programming language with dynamic typing.
  The language offers great flexibility for destructing values through
  pattern matching and dynamic type tests.
  Erlang also comes with a type language supporting parametric polymorphism,
  equi-recursive types, as well as union and a limited form of intersection types.
  However, type signatures only serve as documentation; there is no check
  that a function body conforms to its signature.

  Set-theoretic types and semantic subtyping fit Erlang's feature set
  very well. They allow expressing nearly all constructs of its type language
  and provide means for statically checking
  type signatures. This article brings set-theoretic types to Erlang
  and demonstrates how existing Erlang code can be statically
  type checked without or with only minor modifications to the code.
  Further, the article formalizes the main ingredients of the type system
  in a small core calculus, reports on an implementation of the system, and
  compares it with other static type checkers for Erlang.
\end{abstract}



\maketitle

\section{Introduction}
\label{sec:introduction}

Erlang is a functional programming language with built-in support for
lightweight processes, distribution, and on-the-fly code reloading.
Since its inception in the mid 1980s~\cite{conf/hopl/Armstrong07},
it has gained a great range of users in various industries.\footnote{%
  See \url{https://erlang-companies.org} for an incomplete list of industrial Erlang users.
}
Writing code that runs in production requires extra care when
dealing with erroneous situations. A common approach in Erlang for error handling
is to let the problematic process crash and restart it.

For a large class of errors, this \enquote{let it crash} philosophy works surprisingly
well. However, some errors cannot be rectified by a simple restart. For example,
code that tries to add a number and some string will always fail. The reason
for this error is a misunderstanding by the programmer about data formats.
Static type systems eliminate such errors
before running the program~\cite{TAPL}. Since Erlang is a dynamically typed
language, these errors only appear when running the program.

There have been various attempts to add static type checking to
Erlang~\cite{dialyzer,gradualizer,conf/icfp/MarlowW97,conf/erlang/ValliappanH18,conf/erlang/RajendrakumarB21,conf/coordination/MostrousV11,eqWAlizer}.
Erlang even offers a language to
express types, type definitions, and type signatures (called specifications) of
functions~\cite[\S\,7]{ErlangReference2022}.
However, the Erlang compiler does not check a type signature
against the implementation; the signature merely serves as documentation.

Among the attempts, as mentioned above, to statically type-check Erlang code,
the tool that gained greater adoption is
Dialyzer~\cite{dialyzer}. The Dialyzer tool performs static type checking
based on success typing~\cite{conf/ppdp/LindahlS06}. It reports
an error only if the error will definitely result in a runtime crash.
Hence, even if Dialyzer successfully checks an Erlang program, the program might still
crash because of a data format mismatch.
In contrast, the
type systems of almost any other statically typed language
(Haskell, ML, Java, Rust, to name just a few)
guarantee the absence of certain classes of errors but might emit false positives
in the form of rejecting programs that would run without any crashes.

In this article, we investigate the design of a static type system for Erlang
that offers a greater degree of safety. If the type checker accepts a program,
certain classes of errors never occur at runtime:
argument type mismatch for operations or functions, incomplete pattern
matching, or redundant branches in pattern matching.
At this time of our endeavor, we
restrict ourselves to sequential Erlang code without dynamic code loading
or evaluation of dynamically constructed expressions.
%However, we believe that our
%approach is also extensible to these situations.

Our motivation in designing a static type system for Erlang rests on practical
experience with several large Erlang projects \cite{antidote,riakcorelite}.
We believe that a static type system will improve software quality in
these projects while reducing development and maintenance time in the long term.
Furthermore, the desire to use the type checker for existing Erlang projects
means that it should work on idiomatic Erlang code.
It should be possible to successfully type check
Erlang programs with no or only minor modifications to the source code,
provided these programs have been written \enquote{with types in mind}.
Such programs follow good developer practice: top-level functions come equipped
with a type signature, and important abstractions are documented through type definitions.
From our experience, many Erlang projects follow this good practice.

Several features commonly used in sequential Erlang and its type language
makes static type checking
challenging: dynamic type tests, some form of subtyping, equi-recursive
types~\cite{conf/pldi/CraryHP99,conf/lics/AbadiF96},
polymorphic types, intersection types in signatures, and the ad-hoc formation of new types
through the use of untagged union types~\cite{Pierce1992}.
The standard approach for defining a subtyping relation would be via a set of deduction
rules relating two types based on their \emph{syntactic form}.
However, intersection and union types pose a number of challenges to this approach.
For example, distributivity laws do not hold as expected, and
least upper / greatest lower bounds cannot be expressed in general~\cite{Pierce1992,conf/tacs/Damm94}.

\emph{Semantic
subtyping}~\cite{journals/jacm/FrischCB08,conf/icfp/CastagnaX11,conf/popl/Castagna0XILP14,conf/popl/Castagna0XA15}
takes a different route by interpreting types as subsets of a model, resulting
in \emph{set-theoretic types}.
Subtyping is then simply set inclusion; intersection, union, and even negation types
arise quite naturally as their set-theoretic counterparts.
While this approach is intuitive and elementary, it also has its price: the model of
types must be carefully designed to match the semantics of the language
and the algorithms for deciding subtyping and subtyping constraints are intricate and computationally complex.

However, it turns out that the model for set-theoretic types defined by Castagna and
coworkers~\cite{conf/icfp/CastagnaX11,conf/popl/Castagna0XILP14,conf/popl/Castagna0XA15}
fits Erlang very well. Further, their work on typing polymorphic variants with
set-theoretic types~\cite{conf/icfp/CastagnaP016} provide solutions to some problems in connection
with pattern matching and type inference that arise in a similar form for Erlang.
Thus, the design of our type system rests on set-theoretic types supporting rank-1 polymorphism,
a limited form of type inference (intersection types must be introduced by type signatures),
and type tests but only on the outermost type constructor~\cite{Castagna2021}.
Occurrence typing \cite{conf/popl/Tobin-HochstadtF08,journals/pacmpl/CastagnaLNL22}
refines the type of variables through pattern matching and types tests.
Nearly all features of Erlang's type language are supported, with non-regular polymorphic
types being the only notable omission.

% maybe?:
% mention relation erlang <- -> core erlang
% why we don't care about core erlang

\paragraph{Contributions}
This article introduces \emph{\NAME{}}, a static type checker for Erlang\footnote{\url{https://github.com/etylizer/etylizer}}.
To our knowledge, \NAME{} is the first adaption of set-theoretic types to a
dynamically typed programming language widely used in industry.
Since we were not involved with the research team where the idea and implementations
of set-theoretic types originated, our work can also be considered as some form of
reproduction study. We make the following contributions:

\begin{itemize}
\item We show several examples from Erlang's standard library demonstrating how a type
  system based on set-theoretic types allows to express many idioms of sequential Erlang code
  without changing existing programs.
\item We adapt and extend set-theoretic types to the full Erlang type language.
\item We formalize the essential ingredients of set-theoretic types for Erlang as a small core calculus.
\item We provide an implementation of \NAME{} that covers most of Erlang's language constructs.
\item We evaluate the effectiveness of \NAME{} by comparing it with two other static type checkers
  for Erlang on more than 300 test cases.
%  The resulting tool is called \emph{\NAME{}}.
\end{itemize}

\paragraph{Roadmap} \Cref{sec:examples} introduces \NAME{} through a series of examples,
\Cref{sec:formal-type-system} formalizes its main ideas, and \Cref{sec:implementation}
reports on our implementation and evaluation. \Cref{sec:related-work} discusses related work,
whereas \Cref{sec:future-work-concl} gives pointers to future work and concludes.

%%% Local Variables:
%%% mode: latex
%%% TeX-master: "typing-erlang-ifl-2022.tex"
%%% End:

\section{Examples}
\label{sec:examples}

To introduce the principles of set-theoretic types, we start by discussing some aspects of the Erlang (type) language that are difficult to cover in rule-based type systems, but a good fit for set-theoretic types and semantic subtyping.

\subsection{Singleton, Union, and Intersection types}

As a first example, consider the function \lstinline|last_day_of_the_month/2| from the standard library that determines the number of days in a month.\footnote{\url{https://www.erlang.org/doc/man/calendar.html\#type-ldom}}

\begin{lstlisting}[language=Erlang,numbers=left]
last_day_of_the_month(Y, M) when is_integer(Y), Y >= 0 ->
    last_day_of_the_month1(Y, M).

last_day_of_the_month1(_, 4) -> 30;
last_day_of_the_month1(_, 6) -> 30;
last_day_of_the_month1(_, 9) -> 30;
last_day_of_the_month1(_,11) -> 30;
last_day_of_the_month1(Y, 2) ->
   case is_leap_year(Y) of
      true -> 29;
      _    -> 28
   end;
last_day_of_the_month1(_, M) 
  when is_integer(M), M > 0, M < 13 ->
    31.
\end{lstlisting}

In Erlang, functions consist of at least one function clause. 
A clause consists of a function header and the corresponding function body.
A header comprises the function's name, a sequence of arguments separated by~\lstinline|,| and enclosed in parentheses,
and the arrow symbol~\lstinline|->|, which separates the function header from the function body.
The function body consists of a sequence of expressions separated by~\lstinline|,|.
The body is terminated with the last expression in the sequence by~\lstinline|.|.
The last expression is the value that is returned when the function is evaluated.
In our example, the first parameter~\lstinline|Y| can be any integer larger or equal to zero, while \lstinline|M| is constrained to be an integer larger than 0 and smaller than 13.
The constraints on the parameters are defined with guards in the \lstinline|when| clauses.
The function hence maps values from the set $\{4,6,9,11\}$ to $30$, 2 is mapped to either 29 or 28 and values in $\{1, \dots, 12\} \setminus \{4,6,9,11,2\} = \{1,3,5,7,8,10,12\}$ are mapped to 31.

Type specifications are optional in Erlang.
They can be added to any function with the \lstinline|-spec| annotation
and employ a similar syntax to expressions, but use the Erlang type specification language instead.

The most precise set-theoretic type for the above function -- short of allowing arithmetic operations on type level -- would be:

\begin{lstlisting}[language=Erlang,numbers=none]
-spec last_day_of_the_month
    (non_neg_integer(), 2) -> 28 | 29
  ; (non_neg_integer(), 4 | 6 | 9 | 11 | 2) -> 30;
  ; (non_neg_integer(), 1 | 3 | 5 | 7 | 8 | 10 | 12) -> 31.
\end{lstlisting}

These three function clause types correspond naturally to an intersection of three arrow types.
Though this type is expressible in Erlang's type language, the Erlang standard library provides a less precise type signature for the function:\footnote{The standard library actually introduces some type synonyms that we directly inline here.}

\begin{lstlisting}[language=Erlang,numbers=none]
-spec last_day_of_the_month(non_neg_integer(), 1..12) -> 
  28 | 29 | 30 | 31
\end{lstlisting}

% where 

% \begin{lstlisting}[language=Erlang,numbers=left]
% year() = non_neg_integer()
% month() = 1..12
% ldom() = 28 | 29 | 30 | 31
% \end{lstlisting}

In these specifications, the constraints for the function are described by a range with a dedicated type for non-negative integers, the range type operator \lstinline{..}, and singleton types (here for the integer values), and union types, denoted by type constructor \lstinline{ | }.
% The first parameter \lstinline{year()} can be any integer larger or equal to zero, while \lstinline{month()} is an integer in the range of 1 to 12.
%The return type consists of the union of singleton types for the numerical constants 28 to 31.

Intersection types are a direct consequence of function overloading. The different function clauses have each their respective specification, and the type specification of all clauses is given by the intersection of these specifications, denoted by \lstinline{;}. 
Due to the singleton types, as shown in the examples, such overloading can result in a very detailed description of a function's behavior on the type level.

In \NAME{}, we provide a direct correspondence of these types, using the same constructors, and our type-checker can successfully check the function (with either of the two type signatures), thus allowing programmers to trade readability with preciseness without compromising on safety.

%TODO Generalization to polymorphic maybe data type in section on polymorphism?

% When type checking the function, we derive for each expression some subtype constraints that are then resolved using a tally algorithm \cite{XXX}. 
% TODO: Is this order correct?
% First, the guard for \lstinline{last_day_of_the_month\2} yields the range \lstinline|[1,???]|
% for the first parameter, and thus the constrain??? 
% Next, the function clauses for \lstinline{last_day_of_the_month1\2} are analysed.
% According to Erlang's semantic, the first clause that matches is taken.
% Each non-matchin clause therefore restricts the set of values for the parameter which can be consisely expressed using negation types.\footnote{Erlang's type language does not provide an equivalent type constructor. We therefore do not allow to annotate negation types though they are constructed and used during type check. }
% TODO example !
% Finally, the co- and contra-variance of parameters and result add constraints XXX.
% The tally algorithm then yields as substitution XXX, thus providing a proof that the annotated type is correct.

% TODO A naive implementation of the type checker applied to \lstinline{last_day_of_the_month\2}
% requires the construction of XXXX different sets (see Section \ref{sec:practical-experience} for more details).
% %TODO Is this actually discussed in Section practical-experience??

Another use for singleton types are atoms.
An Erlang atom is a literal that introduces a constant whose value is its own name.
They are typically used for enumerations and in tagged tuples.

% \begin{lstlisting}[language=Erlang,numbers=left]
% -spec sign(integer()) -> plus | minus.
% sign(X) = if 
%     X < 0 -> minus;
%     true  -> plus
% end.
% \end{lstlisting}
    
In the following example, atoms are used to implement a maybe data type.

\begin{lstlisting}[language=Erlang,numbers=left]
-spec safe_div(any(), 0) -> none
  ; (integer(), pos_integer() | neg_integer()) 
      ->  {ok, integer()}
safe_div(X, Y) -> case Y of 
    0 -> none;
    _ -> {ok, X div Y} end.
\end{lstlisting}

The function \lstinline|safe_div/2| yields the result of dividing an integer by another, non-zero integer; otherwise, it returns \lstinline|none|.
As we have seen with integers above, each atom \lstinline|ok| and \lstinline|none| is associated with their individual singleton type.
The example further shows tuple types with their usual semantics and their type constructor \lstinline|{ , }| and the top type \lstinline|any()|, which comprises all values and is hence a superset of all other types.
Similar to the function clauses in the previous example, the analysis of the \lstinline|case| expression in the function body yields an intersection of two different types.

As the Erlang type language does not provide negation types, we decided to also not support them in type signatures for \NAME{}. 
The type for all non-zero integers is here instead described as the union of positive and negative integers.
However, negation types arise when analyzing \lstinline|case| expressions.
In \lstinline|safe_div/2|, the second case clause uses \lstinline|Y| as a guarded expression which restricts the type of \lstinline|Y| to all integers other than 0, since otherwise, the first clause would have applied.
For each of the clauses, \NAME{} further checks that the type of the guarded expression is inhabitable and that the clauses cover all possible values for the expression.

The function further highlights some typical use of tagged unions as result type. 
In contrast to other Erlang typing tools like Dialyzer, \NAME{} can also correctly check the following variant of the function that does not rely on the tagged tuple:

\begin{lstlisting}[language=Erlang,numbers=left]
-spec safe_div_alt(any(), 0) -> none
  ; (integer(), pos_integer() | neg_integer()) 
      -> integer().
safe_div_alt(X, Y) -> case Y of 
    0 -> none;
    _ -> X div Y end.
\end{lstlisting}

%\subsection{Function overloading}

% \begin{lstlisting}[language=Erlang,numbers=left]
% -spec foo(integer()) -> integer(); (1|2) -> (1|2).
% foo(X) -> case X of
%     1 -> 2;
%     2 -> 1;
%     Y -> Y + 1
% end.
% \end{lstlisting}

% TODO: Besser eine boolsche Funktion? Oder ist das zu ähnlich zu obigem Beispiel?
% \begin{lstlisting}[language=Erlang,numbers=left]
% -type light() :: red | yellow | green.
% -spec next(light()) -> light()
% next(Color) = case Color of
%     red    -> green;
%     yellow -> red;
%     green  -> yellow
% end.
% \end{lstlisting}

% TODO: Explain!
% \begin{itemize}
% \item Different types possible:
% \begin{lstlisting}[language=Erlang,numbers=left]
%     -spec next(red | yellow | green) -> red | yellow | green
%     \end{lstlisting}
%     \begin{lstlisting}[language=Erlang,numbers=left]
%         -spec next(red) -> green; yellow -> red; red -> green
%         \end{lstlisting}
% \end{itemize}

\subsection{Recursive data types}

Erlang has the capability to define custom types.
The type to be defined can be used on the right side again to define recursive types.
The following example defines a simple recursive monomorphic tree type for integers.
The idiomatic tagged tuples are used again to define
the two cases of the recursive data type; the leaf nodes and the branching nodes.

\begin{lstlisting}[language=Erlang,numbers=left]
-type tree() :: nil | {node, integer(), tree(), tree()}.

-spec find_node(integer(), tree()) -> boolean().
find_node(_, nil) -> false;
find_node(N, {node,N, Lt, Rt}) -> true;
find_node(N, {node,_, Lt, Rt}) ->
    find_node(N, Lt) andalso find_node(N,Rt).

-spec lookup() -> boolean().
lookup() ->
    BadTree = {node, 3, 
        {node, 1, nil, {nil, bad}}, 
        nil
    },
    find_node(5, BadTree).
\end{lstlisting}

The function \lstinline|find_node/2| walks a given tree and returns true or false depending on whether a given element is contained in the tree or not. 
The function \lstinline|lookup| generates a \lstinline|tree()| type on the fly and tries to apply the \lstinline|find_node/2| function.
Our tool \NAME{} supports recursive types and detects that the tagged tuple \lstinline|BadTree| does not conform to the type specified in \lstinline|tree()| and should thus not be used as parameter in \lstinline|find_node/2|.

\subsection{Polymorphism}

The Erlang type language also supports parametrically polymorphic functions and data types.
The following function is taken with its type signature from Erlang's standard library:

\begin{lstlisting}[language=Erlang,numbers=left]
-spec filtermap(fun((T) -> boolean()), [T]) -> [T]
    ; (fun((T) -> {true, U} | false), [T]) -> [U]
    ; (fun((T) -> {true, U} | boolean()), [T]) -> [T | U].
filtermap(_F, []) -> [];
filtermap(F, [X|XS]) ->
    case F(X) of
        false -> filtermap(F, XS);
        true -> [X | filtermap(F, XS)];
        {true, Y} -> [Y | filtermap(F, XS)]
    end.
\end{lstlisting}

As the name and signature indicate, \lstinline|filtermap/2| filters elements based on some Boolean function or applies an additional transformation on the filtered values (from domain \lstinline|T|) to obtain values from \lstinline|U|, or yields a list mixing transformed and non-filtered values.

Our tool \NAME{} can successfully type-check this higher-order polymorphic and recursive function.
In the implementation, lists are treated as a built-in data type according to their implementation in Erlang.

We can also turn the above tree data type into a polymorphic one. 
We could define the tree type parametrically over its content type:

\begin{lstlisting}[language=Erlang,numbers=none]
-type tree(A) :: nil | {node, A, tree(A), tree(A)}.
\end{lstlisting}

Certain classes of advanced polymorphic data types are not supported, however.
Consider the following example of the \emph{nested datatype} \lstinline|perfect|\cite{conf/mpc/BirdM98, journals/jfp/Hinze00}:

\begin{lstlisting}[language=Erlang,numbers=none]
-type perfect(A) :: A | perfect({A, A}).
\end{lstlisting}

The interesting thing is that on the right side of the equation, \lstinline|perfect({A,A})| is not identical to the left side of the definition.
The type argument \lstinline|A| is wrapped in a tuple and duplicated.
Essentially, \lstinline|perfect(A)| describes complete binary trees where the leaf nodes are of type \lstinline|A|.
Our type system can not handle these types. 
A central assumption of the structure of recursive types is violated, and therefore type definitions like \lstinline|perfect| not allowed in the formalization.

\subsection{Type Tests}

A common idiom of Erlang programs is to use dynamic type tests to execute different parts of code depending on the type at hand.
The following is an excerpt of a pretty printing function that handles the Erlang abstract syntax tree for singleton types.

\begin{lstlisting}[language=Erlang,numbers=left]
-spec ty(integer(), ast:ty()) -> doc().
ty(Precedence, {singleton, A}) ->
  case A of
    _ when is_atom(A) -> text(atom_to_list(A));
    _ when is_integer(A) -> text(integer_to_list(A));
    _ -> text([$$, A]) % must be a char
 end.
\end{lstlisting}

Singleton types in Erlang can either be atoms, integers, or chars.
For each of those three cases, we want to use a different textual representation.
Similar to function clause guards in the first example,
we can employ guards to \lstinline|case| clauses to execute different branches depending on the type of the expression.
Guards are a subset of valid Erlang expressions, and the evaluation of such guards must be guaranteed to be free of side effects.
In \NAME{}, we can statically decide whether each branch is well-typed with the refined guard type.

As the examples show, set-theoretic types are a natural fit for Erlang's type language.
From our (limited) experience, existing type signatures do not need to be altered, and all language constructs from sequential Erlang can be handled to reflect the language semantics.
In particular, type checks and guarded expressions that are idiomatic in Erlang for function clauses and \lstinline|case| expressions can statically be checked for common mistakes that typically lead to semantic errors in the program.

%%% Local Variables:
%%% mode: latex
%%% TeX-master: t
%%% End:

\section{Formal Type System}
\label{sec:formal-type-system}

This section presents a formal foundation for our quest in
devising a type system for Erlang. It defines the language \MinErl{},
a core calculus capturing the
essential ingredients of sequential programming
in Erlang. Our calculus is not a proper subset of Erlang, but
extends the $\lambda$-calculus with
constants, pairs, pattern matching, and mutually recursive
let-bindings. As seen in \Cref*{sec:examples}, pattern matching supports a limited form of
type introspection at runtime. The type system of
\MinErl{} is based on subtyping, polymorphic set-theoretic
types~\cite{journals/jacm/FrischCB08,conf/icfp/CastagnaX11}, and
occurrence typing~\cite{conf/popl/Tobin-HochstadtF08}.
Its formulation is influenced
by the work of Castagna and colleagues on using set-theoretic types
to model polymorphic variants~\cite{conf/icfp/CastagnaP016}.

\subsection{Syntax}

\boxfigSingle{f:syntax}{Syntax of \MinErl}{
  \small
  \[
    \begin{array}{c}
    \begin{array}{r@{\quad}l@{\qquad}r@{\quad}l}
      \textrm{Type variables}       & \Tyvar, \TyvarAux &
      \textrm{Finite sets of type variables} & \TyvarSet\\
      \textrm{Expression variables} &  x, y &
      \textrm{Singleton types (ints, atoms)}            & \Constty
    \end{array}\\
      \textrm{Constants (ints, floats, atoms)}~~\Const
    \end{array}
  \]
  \[
    \begin{array}{rr@{~}r@{~}l@{~}l}
      \textrm{Type schemes} &
      \Polyty & ::= & \TyScm{\TyvarSet} \Monoty
      \\
      \textrm{Mono types} &
      \Monoty  & ::= &
                      \Monoty \Union \Monoty
             \mid \Monoty \Inter \Monoty
             \mid \Neg\Monoty
             \mid \ErlTop
             \mid \ErlBot \\ \textrm{(coinductive)}&&
             \mid &
             \Monoty \to \Monoty
                    \mid \Pairty{\Monoty}{\Monoty}
                    \mid \Tyvar
             \mid \Basety\\
     \textrm{Base types} & \Basety & ::= & \Constty \mid \Int \mid \Float \mid \Atom \mid \AnyPair \mid \AnyFun
      \\[\medskipamount]
      \textrm{Expressions}
         & e & ::= & x
           \mid \Const
           \mid \Abs{x} e
           \mid \App{e}{e}
           \mid \Pair{e}{e} \\ &&
           \mid & \Case{e}{\Multi{\Cls}}
           \mid \Letrec{\Multi{\DefSym}}{e} \\
      \textrm{Pattern clauses}
         & \Cls & ::= & \PatCls{\Pg}{e}                \\
      \textrm{Guarded patterns}
         & \Pg & ::= & \WithGuard{p}{g}                \\
      \textrm{Patterns}
         & p & ::= & \Const \mid \Wildcard \mid x \mid \Pair{p}{p}                 \\
      \textrm{Guards}
         & g & ::= & \Is{b}{x} \mid \GuardAnd{g}{g} \mid \GuardTrue \mid \GuardOracle \\
      \textrm{Definitions}
         & \DefSym  & ::= & \Def{x}{\Polyty}{\Abs{y}{e}} \mid \DefNt{x}{\Abs{y}{e}} \\
    \end{array}
  \]
}

\Cref{f:syntax} defines the syntax of \MinErl{}.
We let $\Tyvar, \TyvarAux$ range over type variables,
$\TyvarSet$ over finite sets of type variables, and
$x, y$ over expression variables.
Metavariable $\Const$ ranges over a
set of constants for ints, floats, and atoms.
The set $\Constty$ contains the
constants that may be used as singleton types; that is,
ints and atoms, but not floats.
%\Todo{What about chars??}
The notation $\mathfrak{s}_{i \in I}$ for some syntactic
construct $\mathfrak{s}$ and some index set
$I = \{1, \ldots, n\}$ is short for the
sequence $\mathfrak{s}_1 \dots \mathfrak{s}_n$.

\subsubsection{Types}

A type scheme $\Polyty$ quantifies a monomorphic type $\Monoty$ over
a finite set of type variables.
At some places, we identify the monomorphic type $\Monoty$ with the
type scheme $\TyScm{\emptyset}{\Monoty}$.
Monomorphic types $\Monoty$ encompass set-theoretic connectives
(the first line in the definition of $\Monoty$, to be explained
shortly), type constructors for functions and pairs, as well as type variables and
base types. Base types $b$ consist of singleton types, types for ints, floats, and atoms, as well
as types $\AnyPair$ and $\AnyFun$ for arbitrary pairs and functions.
For example, the product type $\Pairty{\Int}{\Atom}$ describes
the set of pairs $(i, a)$ where $i$ is an int and $a$ an atom.
In contrast, the type
$\AnyPair$ describes all pairs without restricting the types of the components.

Set-theoretic types require somewhat more explanation.
We only sketch the main ideas here and refer to the literature
for all
details~\cite{journals/jacm/FrischCB08,conf/icfp/CastagnaX11,conf/popl/Castagna0XILP14,conf/popl/Castagna0XA15,conf/icfp/CastagnaP016}.
With set-theoretic types, we interpret types as subsets of some model
\Model{}. Subtyping between types is then just set-inclusion.
Intuitively, \Model{} could be the set of all well-typed values of the
programming language under investigation; a type $t$ then should
denote the set of all values of this type. For technical reasons, such
a definition of \Model{} does not work well in practice. Fortunately,
there are alternative definitions that allow us to recover
the intuitive model~\cite{conf/ppdp/CastagnaF05,journals/jacm/FrischCB08}.

In general, the definition of a model for set-theoretic types is
tied to a specific programming language, in particular to its mode of
evaluation (strict or not) and to the available type constructors.
However, Erlang is close enough to the
languages considered in the aforementioned literature, so the model
and algorithms developed in the work of Castagna et al.~\cite{conf/popl/Castagna0XILP14,conf/popl/Castagna0XA15,conf/icfp/CastagnaP016}
directly transfer to our setting.

Coming back to the syntax of monomorphic types in \MinErl{}
(\Cref{f:syntax}),
the connectives $\Monoty \Union \Monoty'$ and
$\Monoty \Inter \Monoty'$ denote the union and intersection of the
sets associated with $\Monoty$ and $\Monoty'$.
Negation $\Neg \Monoty$ denotes complement with respect to
the model, whereas top type $\ErlTop$ and bottom type
$\ErlBot$ denote the full model and the empty set, respectively.
We employ the convention that all type connectives and constructors associate
to the right, and the order of precedence is $\Union, \Inter, \Neg, \to, \Pairty{}{}$
(ascending).

To allow for recursive types, the definition of monomorphic types $t$
is to be read coinductively. Hence, a monomorphic type is a potential
infinite tree.
For example, the type for lists can be defined as the type
fulfilling the equation
$\ListTy{\Tyvar} = \AtomNil \Union \Pairty{\Tyvar}{\ListTy{\Tyvar}}$. Here, $\AtomNil$ is
a singleton type for the atom $\AtomNil$. The $\ListTy{\Tyvar} =$ is not part
of the type syntax, but only allows us to write down the infinite tree
$\AtomNil \Union \Pairty{\Tyvar}{(\AtomNil \Union \Pairty{\Tyvar}{\ldots})}$
with a finite description.

We require the potentially infinite trees
to be \emph{regular} and \emph{contractive}~\cite{conf/icfp/CastagnaP016, TAPL}.
Regularity requires that a tree has only a finite number of different subtrees.
This condition is crucial for establishing decidability of the subtyping algorithm.
Contractiveness states that every infinite branch has infinitely many occurrences
of the type constructors $\to$ and $\Pairty{}{}$. This condition rules out
nonsensical types fulfilling equations such as $t = t \Union t$ or $t = \Neg t$.

Erlang's type syntax supports union and intersection types, but the latter only
as the top-level connective of a type specification. Our formalization supports
full set-theoretic types because several typing rules require their expressiveness.

\subsubsection{Expressions}

Expressions $e$ include variables $x$, constants $\Const$,
lambda abstractions $\lambda x.e$, function application $e_1\,e_2$, pairs $\Pair{e_1}{e_2}$,
as well as a \kw{case}-expression for pattern matching and
a \kw{letrec} construct for defining mutually recursive functions.
The functions defined by a \kw{letrect} carry an optional type annotation.
By convention,
function application associates to the left and
the body of a lambda abstraction extends as far to the right as possible.
We sometimes use the abbreviation $\lambda x\,y.e$ instead of $\lambda x.\lambda y.e$.

The expression $\Case{e}{\Multi{\Cls}}$ matches scrutinee $e$
against pattern clauses $\Cls_i$. A pattern clause has the form
$\PatCls{\WithGuard{p}{g}}{e'}$, where $p$ is a pattern, $g$ a guard, and
$e'$ the body expression.
Patterns consist of constants $\Const$, wildcards $\Wildcard$,
variables $x$, and pair patterns $\Pair{p}{p}$. At some places, we identify
a pattern $p$ with the guarded pattern $\WithGuard{p}{\GuardTrue}$.

In Erlang, patterns
have a somewhat unusual scoping rule: a pattern variable already
bound in the outer scope only matches if the value being matched is
equal to the variable's value from the outer scope.
In contrast, \MinErl{} uses a more standard approach:
matching against a variable always succeeds and binds
the value being matched to the variable. We took this approach
because its formalization is simpler and the scoping rules for pattern
variables are largely orthogonal to our endeavor.

A guard $g$ further constrains a match. It is essentially a conjunction of type tests against
base types. Such a type test, written $\Is{b}{x}$,
corresponds to guard functions in Erlang such as
\lstinline|is_integer|. To simplify the analysis of type tests,
we only allow variables as arguments. In Erlang, guards may also
contain conditions that are not statically decidable. Such conditions
are summarized in the guard $\GuardOracle$.

\subsection{Dynamic Semantics}

\boxfigSingle{f:dynamic-patterns}{Dynamic semantics of patterns}{
  \[
  \begin{array}{rr@{~}r@{~}ll}
    \textrm{Substitutions} & \Valsubst\\
    \textrm{Values}
       & v & ::= & \Const \mid \Abs{x} e \mid \Pair{v}{v} \\
    \textrm{Extended guards}
       & g & ::= & \ldots \mid \Is{b}{e} \\
  \end{array}
  \]
  \RuleSection{
    \RuleForm{
      \PatSubst{v}{\Pg} = \Valsubst \mid \PatSubstFail \qquad
      \PatSubst{v}{p} = \Valsubst \mid \PatSubstFail
    }
  }{
    Matching of patterns
  }
  \begin{mathpar}
    \inferrule[match-false]{
      \EvalGuard{g} = \False
    }{
      \PatSubst{v}{(\WithGuard{p}{g})} = \PatSubstFail
    }

    \inferrule[match-true]{
      \EvalGuard{g} = \True
    }{
      \PatSubst{v}{(\WithGuard{p}{g})} = \PatSubst{v}{p}
    }

    \inferrule[match-const]{}{
      \PatSubst{v}{\Const} = {
        \begin{cases}
          [\,] & \textrm{if}~ v = \Const\\
          \PatSubstFail & \textrm{otherwise}
        \end{cases}
      }
    }

    \inferrule[match-wild]{}{
      \PatSubst{v}{\Wildcard} = [\,]
    }

    \inferrule[match-var]{}{
      \PatSubst{v}{x} = [v/x]
    }

    \inferrule[match-pair]{}{
      \PatSubst{v}{(p_1, p_2)} = {
        \begin{cases}
          \Valsubst_1 \SubstUnion \Valsubst_2 &
          \parbox[t]{4cm}{
            if $v = (v_1, v_2)$ and $\PatSubst{v_i}{p_i} = \Valsubst_i$\\
            for $i=1,2$ and $\Valsubst_1 \ValEq \Valsubst_2$
          } \\
          \PatSubstFail & \textrm{otherwise}
        \end{cases}
      }
    }
  \end{mathpar}

  \RuleSection{
    \RuleForm{
      \EvalGuard{g}= \True \mid \False \qquad
    }
  }{
    Evaluation of guards
  }
  \begin{mathpar}
    \EvalGuard{\Is{b}{v}} = {
      \begin{cases}
        \True   & \textrm{if}~ v \ValMatches b\\
        \False  & \textrm{otherwise}
      \end{cases}
    }

    \EvalGuard{\GuardOracle} ={} ?

    \EvalGuard{\GuardAnd{g_1}{g_2}} = \EvalGuard{g_1} \textrm{~and~} \EvalGuard{g_2}

    \EvalGuard{\GuardTrue} = \True
  \end{mathpar}

  \RuleSection{
    \RuleForm{
      v \ValEq v \quad
      \Valsubst \ValEq \Valsubst \quad
      v \ValMatches b
    }
  }{
    Equivalence and matching
  }
  \begin{mathpar}
    \inferrule[eq-const]{}{
      \Const \ValEq \Const
    }

    \inferrule[eq-pair]{
      v_1 \ValEq v_1' \\
      v_2 \ValEq v_2'
    }{
      (v_1, v_2) \ValEq (v_1', v_2')
    }

    \inferrule[eq-subst]{
      \Forall{x \in \Dom{\Valsubst_1} \cap \Dom{\Valsubst_2}} \\\\\Valsubst_1(x) \ValEq \Valsubst_2(x)
    }{
      \Valsubst_1 \ValEq \Valsubst_2
    }

    \inferrule[match-const]{}{
      \Const \ValMatches \TyOfConst{\Const}
    }

    \inferrule[match-sub]{
      \Const \ValMatches \Basety'\\\\
      \Basety' \IsSubtyBase \Basety
    }{
      \Const \ValMatches \Basety
    }

    \inferrule[match-pair]{}{
      \Pair{v_1}{v_2} \ValMatches \AnyPair
    }

    \inferrule[match-fun]{}{
      (\Abs{x}{e}) \ValMatches \AnyFun
    }
  \end{mathpar}
}

The formalization of \MinErl{}'s dynamic semantics spreads out over
two figures. \Cref{f:dynamic-patterns} defines auxiliaries for evaluating
pattern matching, and \Cref{f:dynamic-semantics} defines a reduction relation.

We start with \Cref{f:dynamic-patterns}.
Metavariable $\Valsubst = \Multi{[e_i/x_i]}$ denotes the capture-avoiding
substitution of expression variables $x_i$ with expression $e_i$.
We write $e\Valsubst$ for applying substitution $\Valsubst$ to expression $e$,
$\Dom{\Valsubst}$ for
the domain of $\Valsubst$, and $\Valsubst_1 \SubstUnion \Valsubst_2$ for
the left-based union of two substitutions.
The notation $\Valsubst(x)$ retrieves the value for variable $x$, implicitly
assuming $x \in \Dom{\Valsubst}$.
The definition of values $v$ is standard.
In the original syntax (\Cref{f:syntax}), guards are not closed
under substitutions because they allow type tests only on variables.
Therefore, we extend the syntax of guards with type tests $\Is{b}{e}$ on arbitrary
expressions.

The functions $\PatSubst{v}{\Pg}$ and $\PatSubst{v}{p}$ define the semantics of pattern matching
(see also \cite[\S\,2.2]{conf/icfp/CastagnaP016}). If value $v$ matches the guarded pattern $\Pg$
or the pattern $p$, then the result is a substitution $\Valsubst$ for
the variables bound by the pattern. Otherwise, the result is
an error $\PatSubstFail$.
A pair pattern $\Pair{p_1}{p_2}$ might contain the same variable $x$ in
sub-patterns $p_1$ and $p_2$.
In this case, all values matched by $x$ must be equal, expressed
as $\Valsubst_1 \ValEq \Valsubst_2$ in rule \Rule{match-pair}. Equivalence $\ValEq$
for values and substitutions is defined in the lower part of \Cref{f:dynamic-patterns}.

The result of evaluating a pattern guard $g$ is written $\EvalGuard{g}$.
We leave the evaluation of oracles unspecified to stress the fact that the type
system must not depend on its result.
A type test $\Is{b}{v}$ requires checking value $v$
against base type $b$, written $v \ValMatches b$. This relation is also defined
in \Cref{f:dynamic-patterns}. For its definition, we assume a mapping
$\TyOfConst{\Const} = \Basety$ that assigns each constant a base type. Floats
are mapped to \Float{}, ints and atoms to their respective singleton types.
Further, $\IsSubtyBase$ denotes the subtyping relation on base types,
defined as the reflexive, transitive closure of $i \IsSubtyBase \Int$ and
$a \IsSubtyBase \Atom$, where $i$ and $a$ are singleton types for ints and atoms, respectively.

\boxfigSingle{f:dynamic-semantics}{Dynamic semantics of \MinErl}{
  \[
  \begin{array}{rr@{~}r@{~}ll}
    \textrm{Evaluation contexts}
       & \EvalCtx & ::= & \Hole \mid \App{\EvalCtx}e \mid \App{v}{\EvalCtx} \mid
                          \Pair{\EvalCtx}{e} \mid \Pair{v}{\EvalCtx} \\ &&
                          \mid & \Case{\EvalCtx}{\Cls} \\
  \end{array}
  \]
  \RuleSection{
    \RuleForm{
      \Reduce{\FunEnv}{e}{e}
    }
  }{
    Reductions
  }
  \begin{mathpar}
    \inferrule[red-abs]{}{
      \Reduce{\FunEnv}{(\Abs{x}{e}) v}{e[v/x]}
    }

    \inferrule[red-case]{
      \PatSubst{v}{\Pg_j} = \Valsubst' \\
      \Forall{i < j}~\PatSubst{v}{\Pg_i} = \PatSubstFail
    }{
      \Reduce{\FunEnv}{\Case{v}{\Multi{(\Pg_i \to e_i)}}}{e_j\Valsubst'}
    }

    \inferrule[red-var]{
      \FunEnv(x) = v
    }{
      \Reduce{\FunEnv}{x}{v}
    }

    \inferrule[red-letrec]{
      \DefSym_i = (\Def{x_i}{\Polyty_i}{e_i}) ~\textrm{or}~\DefSym_i = (\DefNt{x_i}{e_i})\\
      \FunEnv' = \Multi{[e_i/x_i]} \SubstUnion \FunEnv\\
      \Reduce{\FunEnv'}{e}{e'}
    }{
      \Reduce{\FunEnv}{\Letrec{\Multi{\DefSym}}{e}}{\Letrec{\Multi{\DefSym}}{e'}}
    }

    \inferrule[red-context]{
      \Reduce{\FunEnv}{e_1}{e_2}
    }{
      \Reduce{\Valsubst}{\Plug{\EvalCtx}{e_1}}{\Plug{\EvalCtx}{e_2}}
    }
  \end{mathpar}
}

\Cref{f:dynamic-semantics} defines a standard, call-by-value, small-step reduction relation
for \MinErl{}.
A call-by-value evaluation context $\EvalCtx$ is an expression with a hole $\Hole$ such that
the hole marks the point where the next evaluation step should happen.
We write $\EvalCtx[e]$ to denote the replacement of the hole in $\EvalCtx$ with expression $E$.
The relation $\Reduce{\FunEnv}{e}{e'}$
performs a single reduction step from $e$ to $e'$, where
$\FunEnv$ provides access to the $\LetrecSym$-bound variables.
Rule \Rule{rec-case} reduces a case expression to the body of the first branch that
matches its guarded pattern. Substitutions for letrec-bound variables
are performed lazily to allow for mutually recursive functions. Thus,
rule \Rule{red-var} retrieves the value of a letrec-bound
variable from $\Valsubst$, and rule \Rule{red-letrec}
reduces the body of a \kw{letrec} under an extended substitution~$\Valsubst'$.

\subsection{Static Semantics}

This section formalizes a declarative type system for \MinErl{}, influenced by the
work of Castagna and colleagues~\cite{conf/icfp/CastagnaP016}. We rely on a subtyping relation
between set-theoretic types, written $t_1 \IsSubty t_2$. This relation
has been defined elsewhere~\cite{conf/icfp/CastagnaX11,conf/popl/Castagna0XA15}. Here,
we only note that subtyping between set-theoretic types is decidable. We require
our subtyping relation for base types to be compatible with the full subtyping relation;
that is, $b_1 \IsSubtyBase b_2$ implies $b_1 \IsSubty b_2$.

In the following, a variable environment $\Venv = \{ x_i : \Polyty_i \mid i \in I \}$
is a partial mapping from expression variables to type schemes. We let $\EmptyEnv$
denote the empty type environment,
$\Dom{\Venv}$ the domain of $\Venv$, and
$\Venv(x)$ the lookup of $x$'s type in $\Venv$, implicitly assuming
$x \in \Dom{\Venv}$.
The extension of a variable environment is written $\Venv, x : \Polyty$
where the new binding $x : \Polyty$ hides a potential binding for $x$ in $\Venv$.
Similarly, in the concatenation $\Venv,\Venv'$ bindings in $\Venv'$ take precedence over
those in $\Venv$. We write $\Free{\mathfrak s}$ and $\Bound{\mathfrak s}$ to denote the
variables free and bound, respectively, in some construct $\mathfrak s$.

\subsubsection{Typing Guarded Patterns}

\boxfigSingle{f:aux-pattern-types}{Typing guarded patterns}{
  \RuleSection{
    \RuleForm{
      \PatEnv{\Monoty}{\Pg} = \Venv \qquad
      \PatEnv{\Monoty}{p} = \Venv \qquad
      \Env{g} = \Venv
    }
  }{Pattern environments}
  \begin{mathpar}
    \PatEnv{\Monoty}{(\WithGuard{p}{g})} = (\PatEnv{\Monoty}{p}) \InterEnv \Env{g}

    \PatEnv{\Monoty}{\Const} =
    \PatEnv{\Monoty}{\Wildcard} = \EmptyEnv

    \PatEnv{\Monoty}{x} = \{ x : t \}

    \PatEnv{\Monoty}{(p_1, p_2)} =
    (\PatEnv{\ProjL{\Monoty}}{p_1}) \InterEnv (\PatEnv{\ProjR{\Monoty}}{p_2})

    \Env{\Is{b}{x}} = \{ x : b \}

    \Env{\GuardTrue} = \Env{\GuardOracle} = \{ \}

    \Env{\GuardAnd{g_1}{g_2}} = \Env{g_1} \InterEnv \Env{g_2}
  \end{mathpar}

  \RuleSection{
    \RuleForm{\Venv \InterEnv{} \Venv = \Venv}
  }{
    Intersection of environments
  }
  \begin{mathpar}
    (\Venv_1 \InterEnv \Venv_2)(x) =
    \begin{cases}
      \Venv_1(x)  & \textrm{if}~ x \in \Dom{\Venv_1} ~\textrm{and}~ x \notin \Dom{\Venv_2} \\
      \Venv_2(x)  & \textrm{if}~ x \notin \Dom{\Venv_1} ~\textrm{and}~ x \in \Dom{\Venv_2} \\
      \Venv_1(x) \Inter \Venv_2(x)  & \textrm{if}~ x \in \Dom{\Venv_1} ~\textrm{and}~ x \in \Dom{\Venv_2}
    \end{cases}
  \end{mathpar}

  \RuleSection{
    \RuleForm{
      \PgUpperTy{\Pg}{e} = \Monoty \qquad
      \PgLowerTy{\Pg}{e} = \Monoty
    }
  }{
    Potential and accepting types of patterns
  }
  \begin{mathpar}
    \PgUpperTy{\WithGuard{p}{g}}{e} =
    \PatTy{p}{\Env{g}} \Inter
    \PatTy{\PatFromExp{e}}{\Env{g}}

    \PgLowerTy{\WithGuard{p}{g}}{e} = {
      \begin{cases}
        \PgUpperTy{\WithGuard{p}{g}}{e} &
        \parbox[t]{4cm}{
          if $\Free{g} \subseteq \Bound{p} \cup \Bound{\PatFromExp{e}}$
          and $\GuardOracle \notin g$
        }\\
        \ErlBot & \textrm{otherwise}
      \end{cases}
    }
  \end{mathpar}

  \RuleSection{
    \RuleForm{
      \PatTy{p}{\Venv} = \Monoty
    }
  }{
    Types of patterns
  }
  \begin{mathpar}
    \PatTy{\Const}{\Venv} = \TyOfConst{\Const}

    \PatTy{\Wildcard}{\Venv} = \ErlTop

    \PatTy{x}{\Venv} = {
      \begin{cases}
        \Venv(x) & \textrm{if}~ x \in \Dom{\Venv}\\
        \ErlTop     & \textrm{otherwise}
      \end{cases}
    }

    \PatTy{\Pair{p_1}{p_2}}{\Venv} = \Pairty{\PatTy{p_1}{\Venv}}{\PatTy{p_2}{\Venv}}
  \end{mathpar}

  \RuleSection{
    \RuleForm{
      \PatFromExp{e} = p
    }
  }{
    Expressions as patterns
  }
  \[
  \PatFromExp{e} =
  \begin{cases}
    x  & \textrm{if}~ e = x \\
    (\PatFromExp{e_1}, \PatFromExp{e_2}) & \textrm{if}~ e = (e_1, e_2)\\
    \Wildcard & \textrm{otherwise}
  \end{cases}
  \]
}

\Cref{f:aux-pattern-types} defines several auxiliaries for typing guarded patterns.
The functions $\PatEnv{\Monoty}{\Pg}$ and $\PatEnv{\Monoty}{p}$ compute
an environment $\Venv$ for the variables bound in $\Pg$ and $p$ when being
matched against a value of type $\Monoty$. The function $\Env{g}$ computes
the environment resulting from guard $g$.
The definitions are straightforward except for two situations. (1) If two environments
$\Venv_1$ and $\Venv_2$
contain a type for the same variable, we intersect the two types. Intersection
for environments is written
$\Venv_1 \InterEnv \Venv_2$ and defined in the obvious way, see \Cref{f:aux-pattern-types}.
(2) The definition of $\PatEnv{\Monoty}{(p_1, p_2)}$
uses $\ProjL{\Monoty}$ and $\ProjR{\Monoty}$
to extract the left and right component of a pair type $\Monoty \IsSubty \Pairty{\ErlTop}{\ErlTop}$.
See \cite[App.\, C.2.1]{conf/popl/Castagna0XILP14} for details.

The type $\PgUpperTy{\WithGuard{p}{g}}{e}$ is the \emph{potential type} of
a guarded pattern matched against $e$:
if $e$ matches $\WithGuard{p}{g}$
then $e$ must be of type $\PgUpperTy{\WithGuard{p}{g}}{e}$.
Conversely, $\PgLowerTy{\WithGuard{p}{g}}{e}$ is the \emph{accepting type} of
a guarded pattern matched against $e$:
if $e$ has type  $\PgLowerTy{\WithGuard{p}{g}}{e}$ then
$\WithGuard{p}{g}$ definitely matches $e$.
The definition of
potential and accepting types requires two more
auxiliaries. The type of pattern $p$ with respect to some environment $\Venv$, written
$\PatTy{p}{\Venv} = t$, and the view of expression $e$ as a pattern,
written $\PatFromExp{e} = p$.

We explain
the intuition behind potential and accepting types with examples, first considering
only the potential type. Assume as base types $\Int$ and singleton types
for all int constants.
Now consider the following case-expression, where $x$ and $y$ are variables of unknown type.
$$
\Case{(x, y)}{\PatCls{\WithGuard{(1,z)}{\Is{\Int}{z}}}{\ldots}}
$$
The type of the pattern with respect to the environment of its guard is
$\PatTy{(1, z)}{\Env{\Is{\Int}{z}}} = \Pairty{1}{\Int}$.
This is the same as the potential type
$\PgUpperTy{\WithGuard{(1,z)}{\Is{\Int}{z}}}{(x, y)}$.

%$\PatTy{\PatFromExp{(x, y)}}{\Env{\Is{\Int}{z}}} = \Pairty{\ErlTop}{\ErlTop}$
%and $(\Pairty{1}{\Int}) \Inter (\Pairty{\ErlTop}{\ErlTop}) = \Pairty{1}{\Int}$.

Let us consider a slight variation of the example:
$$
\Case{(x, y)}{\PatCls{\WithGuard{(1,\Wildcard)}{\Is{\Int}{y}}}{\ldots}}
$$
This time, the type test is on $y$, the variable from the scrutinee.
The type of the pattern is now
$\PatTy{(1, \Wildcard)}{\Env{\Is{\Int}{y}}} = \Pairty{1}{\ErlTop}$.
To recover the information loss in the second pair component, we view
the scrutinee expression as a pattern and compute its type
$\PatTy{\PatFromExp{(x, y)}}{\Env{\Is{\Int}{y}}} = \Pairty{\ErlTop}{\Int}$.
Intersecting the two types yields the potential type
$\PgUpperTy{\WithGuard{(1,\Wildcard)}{\Is{\Int}{y}}}{(x, y)} = \Pairty{1}{\Int}$.

In the two examples seen so far, the potential and accepting types of
a guarded pattern coincide: the guarded pattern matches if and only if
the scrutinee has type $\Pairty{1}{\Int}$. The situation changes with $\GuardOracle$.
Remember that $\GuardOracle$ represents those parts of a guard that cannot
be analyzed by the type checker (e.g., because it would require to evaluate a
non-terminating expression). Our example now looks like this:
$$
\Case{(x, y)}{\PatCls{\WithGuard{(1,\Wildcard)}{\GuardAnd{\Is{\Int}{y}}{\GuardOracle}}}{\ldots}}
$$
The potential type is still $\Pairty{1}{\Int}$ because
a match implies that the scrutinee has this type.
But the accepting type is $\ErlBot$ as a match on a scrutinee of type
$\Pairty{1}{\Int}$ could still fail if the oracle evaluates to false.
Oracles are not the only reason why the accepting type of a pattern might becomes
$\ErlBot$. In the definition of $\PgLowerTy{\WithGuard{p}{g}}{e}$,
we also resort to $\ErlBot$ if a type test constrains a variable not bound
in $p$ or $\PatFromExp{e}$.

To justify our restrictions for computing the accepting type, we examined several real-world Erlang programs. 
Among those examined were the open source projects Antidote\cite{antidote} with $\approx$12K lines of code and riak\_core\_lite\cite{riakcorelite} with $\approx$15K lines of code.
We found several examples containing type tests on variables bound by a case or occurring in the scrutinee. 
We did not find any example where a type test constrains a non-variable expression or a variable not contained in the scrutinee or bound by the case. 
Clearly, our findings do not imply that such examples do not exist, but they seem to be rare.

\subsubsection{Typing Expressions}

\boxfigSingle{f:decl-typing}{Typing rules for \MinErl}{
  \RuleSection{
    \RuleForm{
      \ExpTy{\Venv}{e}{\Monoty}
    }
  }{
    Typing expressions
  }
  \begin{mathpar}
    \inferrule[d-var]{
      \Monoty \in \Inst{\Venv(x)}
    }{
      \ExpTy{\Venv}{x}{\Monoty}
    }

    \inferrule[d-const]{}{
      \ExpTy{\Venv}{\Const}{\TyOfConst{\Const}}
    }

    \inferrule[d-abs]{
      \ExpTy{\Venv, x:\Monoty_1}{e}{\Monoty_2}
    }{
      \ExpTy{\Venv}{\Abs{x} e}{\Monoty_1 \to \Monoty_2}
    }

    \inferrule[d-app]{
      \ExpTy{\Venv}{e_1}{\Monoty' \to \Monoty}\\
      \ExpTy{\Venv}{e_2}{\Monoty'}
    }{
      \ExpTy{\Venv}{\App{e_1}{e_2}}{\Monoty}
    }

    \inferrule[d-pair]{
      \ExpTy{\Venv}{e_i}{\Monoty_i}\quad(i=1,2)
    }{
      \ExpTy{\Venv}{\Pair{e_1}{e_2}}{\Pairty{\Monoty_1}{\Monoty_2}}
    }

    \inferrule[d-case]{
      \ExpTy{\Venv}{e}{\Monoty}\\
      \Monoty \IsSubty \UnionBig_{i \in I} \PgLowerTy{\Pg_i}{e}\\
      \Forall{i \in I}\quad
      \Monoty_i = (\Monoty \WithoutTy \UnionBig_{j < i} \PgLowerTy{\Pg_j}{e}) \Inter \PgUpperTy{\Pg_i}{e}\\
      {
        \begin{cases}
          \Monoty_i' = \ErlBot & \textrm{if}~ \Monoty_i \IsSubty \ErlBot
          \\
          \ExpTy{\Venv
            \InterEnv (\PatEnv{\Monoty_i}{\PatFromExp{e}})
            \InterEnv (\PatEnv{\Monoty_i}{\Pg_i})
          }{e_i}{\Monoty_i'}
          & \textrm{otherwise}
        \end{cases}
      }
    }{
      \ExpTy{\Venv}{\Case{e}\Multi{(\Pg_i \to e_i)}}{\UnionBig_{i \in I}{\Monoty_i'}}
    }

    \inferrule[d-letrec]{
      \Forall{i \in I}~\Envs{\Venv}{\DefSym_i} = \MetaPair{\Venv_i'}{\Venv_i''}\\
      \DefOk{\Venv, \Venv'_{i \in I}}{\DefSym_i}\\
      \ExpTy{\Venv, \Venv''_{i \in I}}{e}{\Monoty}
    }{
      \ExpTy{\Venv}{\Letrec{\Multi{\DefSym}}{e}}{\Monoty}
    }

    \inferrule[d-sub]{
      \ExpTy{\Venv}{e}{\Monoty'}\\\\
      \Monoty' \IsSubty \Monoty
    }{
      \ExpTy{\Venv}{e}{\Monoty}
    }
  \end{mathpar}

  \RuleSection{
    \RuleForm{
      \DefOk{\Venv}{\DefSym}
    }
  }{
    Typing definitions
  }
  \begin{mathpar}
    \inferrule[d-def-no-annot]{
      \Venv(x) = \Monoty\\
      \ExpTy{\Venv}{\Abs{y}{e}}{t}
    }{
      \DefOk{\Venv}{\DefNt{x}{\Abs{y}{e}}}
    }

    \inferrule[d-def-annot]{
      \FreeTyVars{\Polyty} = \emptyset\\\\
      \Polyty = \TyScm{\TyvarSet}{\InterBig_{i \in I}{(\Monoty_i' \to \Monoty_i)}}\\\\
      \Forall{i \in I}\quad
      \ExpTy{\Venv, x:\Monoty_i'}{e}{\Monoty_i}
    }{
      \DefOk{\Venv}{\Def{x}{\Polyty}{\Abs{y}{e}}}
    }
  \end{mathpar}

  \RuleSection{
    \RuleForm{
      \Envs{\Venv}{\DefSym}
    }
  }{
    Environments from definitions
  }
  \begin{mathpar}
    \inferrule[d-envs-no-annot]{
      \TyvarSet = \FreeTyVars{\Monoty} \setminus \FreeTyVars{\Venv}
    }{
      \Envs{\Venv}{\DefNt{x}{e}} = \MetaPair{\{x : \Monoty\}}{\{x : \TyScm{\TyvarSet}{t}\}}
    }

    \inferrule[d-envs-annot]{}{
      \Envs{\Venv}{\Def{x}{\Polyty}{e}}\\\\ = \MetaPair{\{ x : \Polyty \}}{\{x : \Polyty\}}
    }
  \end{mathpar}
}

\Cref{f:decl-typing} shows the typing rules for \MinErl{}.
Judgments $\ExpTy{\Venv}{e}{\Monoty}$ asserts that under environment $\Venv$ expression $e$
has type $\Monoty$. The rules for
variables, constants, abstractions, applications, pairs, and the subsumption rule \Rule{d-sub}
are standard. In rule \Rule{d-var}, $\InstSym$ denotes the set of monomorphic instances of
some type scheme:
$$
    \Inst{\TyScm{\TyvarSet}{\Monoty}} =
    \{ \Monoty\Tysubst \mid \Dom{\Tysubst} = \TyvarSet \}
$$
We let $\Tysubst = \Multi{[\Monoty_i/\Tyvar_i]}$ denote the type substitution
of type variables $\Tyvar_i$ with monotypes $\Monoty_i$.
We write $\Monoty\Tysubst$ for the application of $\Tysubst$ to some type $\Monoty$, and
$\Dom{\Tysubst}$ for its domain.

The rules for \kw{case} and \kw{letrec} requires more explanation. We start with rule \Rule{d-letrec}.
Type inference with polymorphic recursion is undecidable~\cite{journals/toplas/Henglein93}, so calling a function
recursively with a polymorphic type requires a type annotation.
Hence, the rule first computes two environments from each definition $\DefSym_i$.
Environment $\Venv_i'$ with monomorphic types for functions without
type annotations, $\Venv_i''$ with polymorphic types.
The rule then uses the $\Venv_i'$ for checking the definitions $\DefSym_i$
and the $\Venv_i''$ for typing the body of the letrec.

The auxiliary judgment $\DefOk{\Venv}{\DefSym}$ checks correctness of definition $\DefSym$
under environment $\Venv$. Type inference with intersection types is undecidable in general~\cite{journals/tcs/Bakel95},
so our system never guesses such types. Consequently, rule \Rule{d-def-no-annot} for definitions
without type annotations simply checks that right-hand side of the definition
has the expected type from the environment (i.e., the type guessed previously in rule
\Rule{d-envs-no-annot}). Rule \Rule{d-def-annot} supports
intersection types in annotations by checking the function body against each type
of the intersection. For simplicity, the type scheme $\Polyty$ must be closed. This is no restriction
in practice as Erlang only supports type annotations for top-level functions.

Rule \Rule{d-case} types \kw{case}-expressions. The first premise assigns type $t$ to the scrutinee $e$.
The second premise checks that the branches of the case are exhaustive: $t$ has to be a subtype
of the union of the accepting types of all branches.
The third premise then computes the input type $t_i$ for each branch. This input type is the refined type
of the scrutinee when matching the $i$-th branch. To compute $t_i$,
we subtract the accepting types of all preceding
branches from $t$ as these branches could not have been matched when considering the $i$-th branch. Then
we intersect the result with the potential type of the branch because matching
implies that $t_i$ is some subtype of the potential type.

The last premise checks the body of the $i$-th branch by assigning it the output type $t_i'$.
The type of the whole \kw{case}-expression is then the union of all output types.
If input type $t_i$ is bottom,
then the branch can never match, so the output type $t_i'$ is also bottom.
When typing a function against an intersection type,
skipping unmatched branches is essential. Otherwise, type errors in some branch
would be reported, although the branch can never match for a specific part of the intersection,
see below for an example.\footnote{%
  An implementation should report an error if it detects that a branch does not match for
  all components of an intersection or if there is no intersection at all.}
If input type $t_i$ is not bottom, then we type the body $e_i$ of the branch in an extended and/or refined environment.
This environment contains new bindings for variables introduced by the pattern of the branch, as well as refined
types for variables with type tests in its guard.

\subsubsection{Example}
We conclude the explanation of the typing rules of \MinErl{} by sketching how typing
proceeds for the \lstinline{filtermap} example from \Cref{sec:examples}. Specifically,
we explain how the \kw{case}-expression in the body of \lstinline{filtermap}
type checks against the second component of its intersection type. Here are the relevant parts of the code
in \MinErl{}. Remember that lists are represented as nested pairs with atom
\AtomNil{} as terminator. Atoms \AtomTrue{} and \AtomFalse{} act as Booleans.

\noindent
\scalebox{0.9}{%
$$
\begin{array}[t]{@{}l@{}}
\kw{letrec}\\\phantom{x}
  \Filtermap \\\phantom{x}
  \begin{array}[t]{@{~}r@{~}l@{}}
    :& \forall \Tyvar, \TyvarAux .~
      (\ldots \Inter
      (\Tyvar \to (\Pairty{\AtomTrue}{\TyvarAux} \Union \AtomFalse)) \to \ListTy{\Tyvar} \to \ListTy{\TyvarAux}
      \Inter \ldots)
  \\
  =& \lambda f\, l.\,
  \kw{case}~{l}~\kw{of}~
  {
  \begin{array}[t]{@{}l@{}}
    \AtomNil \to \AtomNil\\
    (x, l') \to~\\\quad
    {
    \begin{array}[t]{@{}l@{}}
      \kw{case}~{f~x}~\kw{of}~
      {
      \begin{array}[t]{@{}l@{}}
        \AtomFalse \to \Filtermap~f~l'\\
        \AtomTrue \to (x, \Filtermap~f~l')\\
        (\AtomTrue, y) \to (y, \Filtermap~f~l')
      \end{array}
      }
    \end{array}
    }
  \end{array}
  }
  \end{array}
\\\kw{in}\,\ldots
\end{array}
$$
}

We show how rule \Rule{d-case} type checks the inner case expression
in environment
$\Venv = \{ f : \Tyvar \to (\Pairty{\AtomTrue}{\TyvarAux} \Union \AtomFalse), l : \ListTy{\Tyvar},
 x : \Tyvar, l' : \ListTy{\Tyvar} \}$.
Scrutinee $e = f~x$ has type $t = \Pairty{\AtomTrue}{\TyvarAux} \Union \AtomFalse$.
The potential and accepting types of the three (guarded) patterns are:\footnote{%
  By slight abuse of notation, we use the $=$ symbol not only for syntactic equality
  between types but also for equivalence module subtyping. For example,
  expanding the definition of $\PgUpperTy{(\AtomTrue, y)}{e}$ yields the type
  $(\Pairty{\AtomTrue}{\ErlTop}) \Inter \ErlTop$, which is equivalent to
  $\Pairty{\AtomTrue}{\ErlTop}$.
}
\[
\begin{array}{r@{~}l@{~}l@{~}l}
\PgLowerTy{\Pg_1}{e} & = \PgUpperTy{\Pg_1}{e} & = \PgUpperTy{\AtomFalse}{e} & = \AtomFalse \\
\PgLowerTy{\Pg_2}{e} & = \PgUpperTy{\Pg_2}{e} & = \PgUpperTy{\AtomTrue}{e}  & = \AtomTrue \\
\PgLowerTy{\Pg_3}{e} & = \PgUpperTy{\Pg_3}{e} & = \PgUpperTy{(\AtomTrue, y)}{e} & = \Pairty{\AtomTrue}{\ErlTop}
\end{array}
\]
Thus, $t \IsSubty \AtomFalse \Union \AtomTrue \Union \Pairty{\AtomTrue}{\ErlTop}$,
so the branches of the case are exhaustive. The input types of the branches are as follows:
\[
\begin{array}{r@{~}l@{~}l@{~}l}
  t_1 & = t \Inter \AtomFalse & = \AtomFalse\\
  t_2 & = (t \WithoutTy \AtomFalse) \Inter \AtomTrue & = \ErlBot\\
  t_3 & = (t \WithoutTy (\AtomFalse \Union \AtomTrue)) \Inter \Pairty{\AtomTrue}{\ErlBot} &= \Pairty{\AtomTrue}{\TyvarAux}
\end{array}
\]
We see that $t_2 = \ErlBot$, which reflects our intuition that the second branch
can never be taken when type checking against the second component of the intersection.
Thus, output type $t_2' = \ErlBot$. For the two other output types, we have
$t_1' = \ListTy{\TyvarAux} = t_3'$. Here, the third branch is checked under the extended
environment $\Venv,y:\TyvarAux$.
Finally, we have $\ListTy{\TyvarAux} \Union \ErlBot \Union \ListTy{\TyvarAux} = \ListTy{\TyvarAux}$ as
type for the whole \kw{case}-expression.

\subsection{Formal Properties and Typing Algorithm}

Type soundness is an important property of any type system.
Our system builds on the type system for polymorphic variants by Castagna and
colleagues~\cite{conf/icfp/CastagnaP016},
which is known to be sound. Thus, we believe
that our system is sound too, but we do not have a proof.

The typing rules in \Cref{f:decl-typing} do not allow to read off a typing algorithm
because the rules are not syntax-directed. The culprit is rule \Rule{d-sub}, which can be
applied on any expression form to lift its type to some supertype.
Thus, we reformulate the typing rules
as a syntax-directed set of contraint-generation rules, again following the work of Castagna and
colleagues~\cite{conf/icfp/CastagnaP016}. The resulting constraints are
then simplified, arriving at a set of subtyping constraints.
Finally, the tally algorithm~\cite{conf/popl/Castagna0XA15} solves
the set of subtyping constraints.

We define the constraint generation and
simplification rules in \Cref{sec:algor-typing-rules}. 
Verifying that the typing rules in \Cref{f:decl-typing} are equivalent to the algorithm obtained by generating, simplifying and solving constraints is future work.

%%% Local Variables:
%%% mode: latex
%%% TeX-master: "typing-erlang-ifl-2022.tex"
%%% End:

\section{Implementation}
\label{sec:implementation}

To validate our formalism, we have implemented a prototype for the rules presented in \cref{sec:formal-type-system}.
In this section, we give an overview of the architecture of the prototype,
the extensions we implemented to cover full Erlang and the challenges of using a semantic approach for type checking.
The implementation employs a constraint solver as used in the type checking of other (functional) languages~\cite{ATTAPL/typeinference}.
The function \lstinline|chk|, shown below, takes as argument a path to a module
and verifies that all exported and annotated functions are well-typed.

\begin{lstlisting}[language=Erlang,numbers=left]
-spec chk(Path) -> ok | throw(err) when Path :: atom().
chk(Path) ->
  RawForms = parse(Path),
  Forms = ast_transform(RawForms),
  Context = default_symtab_extend(Forms),
  lists:foreach(fun(Fun, Spec) ->
    chkfun(Fun, Spec, Context)
  end, Forms).
\end{lstlisting}

Since our prototype is implemented in Erlang, we can make use of the
parsing utilities the language's standard library provides.
In Line 3, the raw abstract syntax tree is returned from the parser.
We use our own sanitized AST format; thus, we next transform the Erlang AST
into our own format in Line 4.
This transformation is only syntactical in nature, and only minor simplifications are employed.

Any expression can evaluate a function or operator that is provided by
the standard Erlang library (i.e., by `erlang.erl`).
To type check a module, we need a context that maintains information about those functions and operators.
The operators are fixed and are defined informally in the Erlang type-language specification.
The built-in functions are not fixed, as the implementation of `erlang.erl` is likely to be extended in the future.
We parse `erlang.erl` once and assume these functions as known in all phases of the type-checking algorithm.
The default functions and operators are stored in a symbol table created in Line~5.
In addition to the globally known signatures, Line~5 extends the table with the local context of the whole module.
Finally, each function with a type annotation is checked with the extended context of the module in Lines~6-8.
The following method checks a function body against its type signature.

\begin{lstlisting}[language=Erlang,numbers=left]
chkfun(Fun, Spec, Context) ->
  MonoTys = mono_ty(Spec),
  lists:foreach(fun(Ty) ->
    chkinter(Fun, Ty, Context)
  end, MonoTys).
\end{lstlisting}

To type check a function, we first transform the potentially polymorphic type signature into a monomorphic one.
Universal quantification of type variables can only occur in the outermost scope of a term in Erlang,
and polymorphic type variables can only have an upper bound.
By removing the universal quantifier and intersecting each type variable with its bound,
we generate the monomorphic version
of the type signature in Line~2.
Since a function signature can be an intersection of multiple types,
we check the function against each of these types in Lines~3-5.

The final part generates the constraints for each intersection type and solves the constraints with
the tally algorithm~\cite{conf/popl/Castagna0XA15}.

\begin{lstlisting}[language=Erlang,numbers=left]
chkinter(Fun, Ty, Context) ->
  RawCs = constr_gen(Fun, Ty),
  SimpCs = const_simp(Ctx, RawCs),
  lists:foreach(fun(Constraint) ->
    case tally(Constraint, Context) of
      [] -> throw(error);
      _ -> ok
  end, SimpCs).
\end{lstlisting}

First, we generate constraints in Line 2
and simplify them in Line~3 (see Figs.~\ref{f:constr-gen} and~\ref{f:constr-rew} in the \cref{sec:algor-typing-rules}).
Each simplified constraint set is then solved with the tally algorithm.
For given types $s$ and $t$, tally finds all substitutions $\theta$ such that $s\theta \IsSubty t\theta$.
If tally returns an empty list of substitutions, the function violates the signature.
Otherwise, the function is well-typed according to its signature.
Since we only perform type checking in our prototype, the solutions are not reused for type inference.
The implementation of tally, which includes the subtyping algorithm, is more involved.
We refer to the work of Castagna and co-workers~\cite{conf/icfp/CastagnaX11, conf/popl/Castagna0XA15} for details,
as we tried to stay as close as possible to the algorithm defined in their articles.

Currently, our implementation requires all code to be placed in one module.
To be practically applicable to large projects,
we will extend the implementation to type check multiple modules.
Such an extension will require only modest engineering effort because we demand that
all exported functions carry a type signature.
The formal system in \Cref{sec:formal-type-system} also supports functions
without signatures. Extending the implementation with type inference
is also on our todo-list, but we will only infer types for module-local functions.

In \cref{sec:formal-type-system} we mentioned that Erlang is (surprisingly) similar to the languages considered in the existing literature on set-theoretic types.
Indeed, we were able to encode the types used in Erlang's type language into the core language defined in prior work~\cite{conf/popl/Castagna0XA15}.
However, we had to discard the encoding approach for our implementation
because of performance reasons and bad error messages. Instead, the implementation
models all Erlang constructs natively.

Nevertheless, encoding can help us understand why Erlang is a good fit for set-theoretic types.
\Cref{f:encoding} shows how the different Erlang types are implemented in terms of equivalent constructs mentioned in \cite{conf/popl/Castagna0XA15}.
The transformation between Erlang types is denoted by $\leadsto$.
Equivalence of concepts from Erlang types to set-theoretic types is denoted by $\equiv_\mathcal{E}$.
Singleton types and atoms are each mapped into their own set corresponding to a set containing basic types.
These two sets are disjoint.
Integers are mapped to integer ranges.
Lists in Erlang are a native language construct and are not defined in terms of a recursive type.
Erlang has two types of lists: proper and improper lists.
Proper lists are always terminated with the special symbol $[]$\footnote{The empty list symbol $[]$ is \emph{not} the atom denoted by '[]'.}.
Improper lists are parameterized over the terminating symbol in addition to  the list's element type.
Since proper lists can be included in improper lists, we define lists in terms of improper lists.
Improper lists are then treated as a product with two covariant type arguments.
We excluded the termination symbol from being included in the set of improper lists as this would imply that everything can be treated as some instance of a type of an improper list.
Erlang's multiple arity functions are transformed into functions with only one argument tuple.
Finally, all tuples are encoded as tuples with arity two.
This encoding ensures that we can treat tuples as set-theoretic products with covariance in both type arguments.

\boxfigSingle{f:encoding}{Encodings and transformations of Erlang types.}{
 \begin{tabular}{lll}
$\text{list } A$ & $\leadsto$ &   (\text{improper\_list } $A$ $[]$) $ \cup$ $[]$   \\
$\{T_1,...,T_i\}$ & $\leadsto$ &   $\{i, \{T_1, \{ ...\{T_{i-1}, T_i\}\}\}\}$ \\
$\text{fun}(T_1,\dots,T_i) \rightarrow B$& $\leadsto$ &   $\text{fun}(\{T_1,\dots,T_i\}) \rightarrow B$ \\
\text{improper\_list } $A$ $B$ & $\equiv_\mathcal{E}$ &   product $(A,B)$ \\
$\{A,B\}$& $\equiv_\mathcal{E}$ &   product $(A,B)$ \\
$\text{fun}(A) \rightarrow B$ & $\equiv_\mathcal{E}$ &   arrow $A \rightarrow B$ \\
atoms & $\equiv_\mathcal{E}$ &  basic types  \\
pid, port, float, ref, $[]$ & $\equiv_\mathcal{E}$ &  basic types \\
integers & $\equiv_\mathcal{E}$ &  integer ranges
\end{tabular}
}

As mentioned before, our implementation does not rely on encodings of native Erlang constructs,
but extends the set-theoretic type language to handle them.
The implemention entails the following steps, which are the main deviations and extensions of Castagna et al.'s work:
\begin{itemize}
	\item subtyping: set-theoretic decomposition with appropriate elimination of top-level variables (see \cite{conf/icfp/CastagnaX11}),
	\item extension of the tally algorithm, defining correct normalization rules (see \cite{conf/popl/Castagna0XA15}), and
	\item extension of the constraint generation algorithm defined in \Cref{sec:algor-typing-rules}.
\end{itemize}

\definecolor{pass}{RGB}{94,176,255} % 5eb0ff
\definecolor{fail}{RGB}{228,48,52}
\definecolor{crash}{RGB}{255,227,80}
\subsection{Capability Evaluation}

\boxfigSingle{f:performance}{Capability Cross-Testing.}{

\pgfdeclareplotmark{etysuite}
{%
	\path[gray, fill=white,postaction={pattern=north west lines}] (-1.5\pgfplotmarksize,-3.8\pgfplotmarksize) rectangle (1.5\pgfplotmarksize,0\pgfplotmarksize);
}
\pgfdeclareplotmark{gradsuite}
{%
	\path[gray, fill=white,postaction={pattern=north east lines}] (-1.5\pgfplotmarksize,-3.8\pgfplotmarksize) rectangle (1.5\pgfplotmarksize,0\pgfplotmarksize);
}

\pgfplotsset{every tick label/.append style={font=\scriptsize}}

\hspace{6mm} % center pushes it too far to the right
\begin{tikzpicture}
	\begin{axis}[
		scale=1.0,
		bar width=15pt,
		ybar stacked,
		nodes near coords custom/.style={
			large value/.style={
			},
			small value/.style={
				/pgfplots/scatter/position=absolute,
				at={(axis cs:\pgfkeysvalueof{/data point/x},\pgfkeysvalueof{/data point/y})},
				text opacity=1,
				inner ysep=1.5pt,
				anchor=south,
			},
			every node near coord/.style={
				check for zero/.code={%
					\pgfmathfloatifflags{\pgfplotspointmeta}{0}{%
						\pgfkeys{/tikz/coordinate}%
					}{%
						\begingroup
						\pgfkeys{/pgf/fpu}%
						\pgfmathparse{\pgfplotspointmeta<#1}%
						\global\let\result=\pgfmathresult
						\endgroup
						\pgfmathfloatcreate{1}{1.0}{0}%
						\let\ONE=\pgfmathresult
						\ifx\result\ONE
						\pgfkeysalso{/pgfplots/small value}%
						\else
						\pgfkeysalso{/pgfplots/large value}%
						\fi
					}
				},
				check for zero,
				font=\small,
			},
		},
		nodes near coords={\pgfmathprintnumber[precision=1]{\pgfplotspointmeta}},
		nodes near coords custom=8,
		% -----------------------------------------------------------------
		legend style={at={(0.5,1.1)},
			anchor=north, legend columns=-1, font=\scriptsize},
		ylabel={\#tests},
		grid,
		xtick=data,
		x tick label style={rotate=45,anchor=east},
		xticklabels={Etylizer, Gradualzier, Dialyzer,Etylizer,Gradualzier,Dialyzer},
		typeset ticklabels with strut,
		% -----------------------------------------------------------------
		ymin=0,
		]

		\addplot+[ybar,black,fill=pass] plot coordinates    {(1,110)   (2,107) (3, 101) (4,68)      (5,166)  (6, 149)};
		\addplot+[ybar,black, fill=fail] plot coordinates      {(1,0)       (2,13)   (3, 24)    (4,3)       (5,30)     (6, 47)};
		\addplot+[ybar,black, fill=crash] plot coordinates {(1,15)     (2,5)      (3, 0)      (4,125)  (5,0)        (6, 0)};

		\fill[green,opacity=0.2,pattern=north west lines] ({rel axis cs:0,0}) rectangle ({rel axis cs:0.5,1});
		\fill[red,opacity=0.5,pattern=north east lines] ({rel axis cs:0.5,0}) rectangle ({rel axis cs:1,1});

		\legend{\strut passed, \strut failed, \strut error}
		\addlegendentry{Etylizer Suite}
		\addlegendimage{opacity=0.2, mark=etysuite}
		\addlegendentry{Gradualizer Suite}
		\addlegendimage{opacity=0.5,mark=gradsuite}

	\end{axis}
\end{tikzpicture}

\begin{center}
\includegraphics[width=0.8\textwidth]{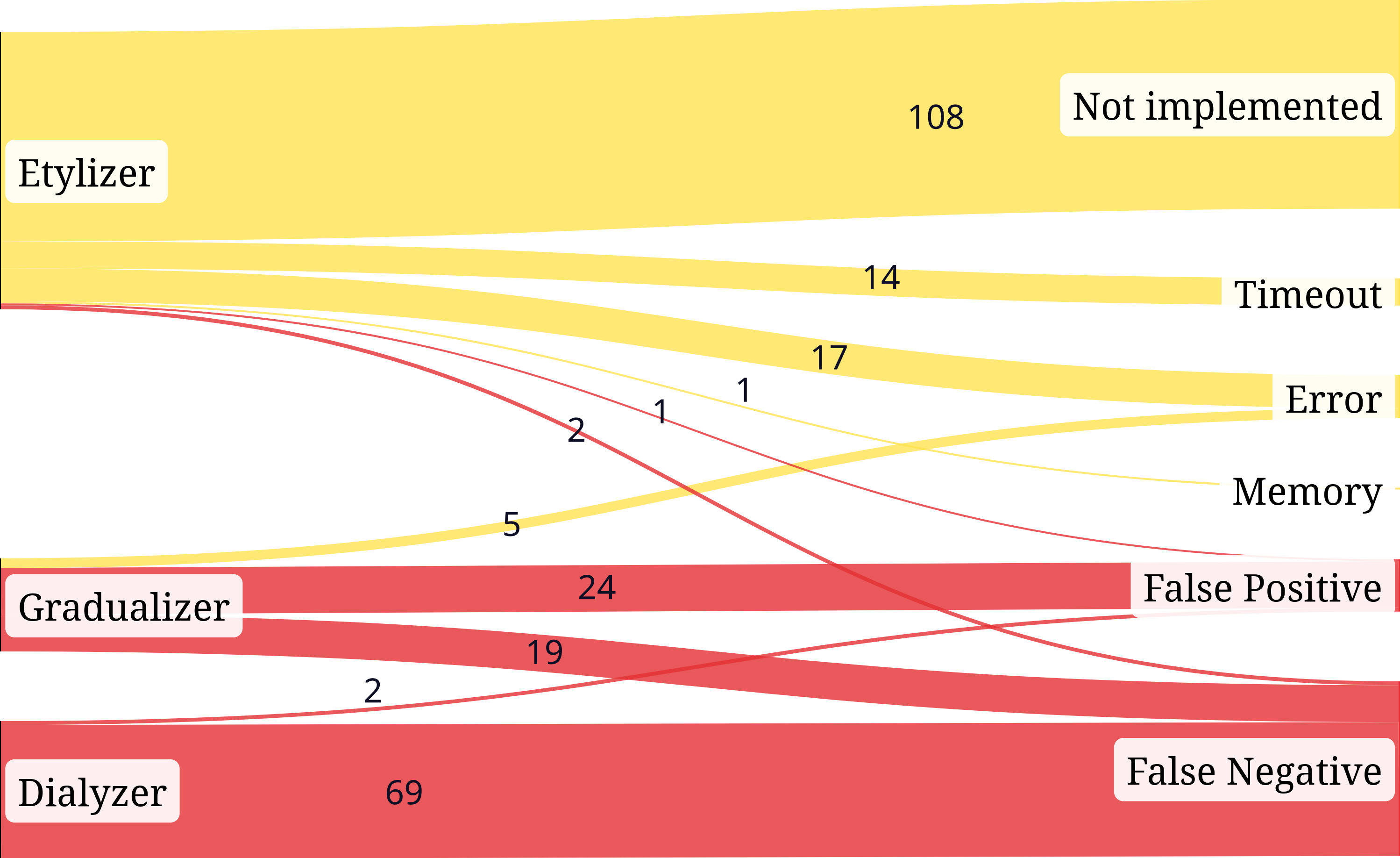}
\end{center}

 }

We evaluated our tool on two benchmark sets which cover a broad set of Erlang's capabilities\footnote{The benchmarks are reproducible and can be found at \url{https://github.com/etylizer/feature-matrix}.}.
The evaluation compares three static type checkers for Erlang: Dialyzer\cite{dialyzer},
Gradualizer\cite{gradualizer}, and our tool \NAME{}.
For the comparison, we took tests from both our test suite and the Gradualizer test suite with
a total of 321 test cases---156 test cases contain known type errors\footnote{We did not include the Dialyzer test suite as it works on a global per-project level and has its own way to distinguish between projects that should fail or pass type checking.}.
Each test case is a single module that is expected to either pass or fail type checking.
We removed tests that depend on definitions external to this module.
Thus we only compare language features, including functionality provided by the built-in \lstinline|erlang| module.
Furthermore, tests related to type inference and tracking precise types for arithmetic expressions were excluded;
these extensions are work in progress for \NAME{}.%
\footnote{An example for tracking precise types for arithmetic expressions is assigning the type \texttt{1|2} to the expression \texttt{X+1} given that \texttt{X} has type \texttt{0|1}.}
%Extending Etylizer with type inference and type-level arithmetic is a work in progress.

\cref{f:performance} shows the test suites divided into the left \NAME{} suite and right Gradualizer suite.
We used a 2.9Ghz quad-core \lstinline|i5-3470s| CPU, limited resident set size memory at 6GB, and time at 10 seconds per module.
Each test suite consists of several test cases that are either type-correct (and should pass the type check) or have some type errors (and should fail the check).
Hence, running a test case with a particular tool may either terminate with the expected result (passed, blue)
or with the wrong result (failed, red), or it may exhibit a runtime error (yellow).
The stacked bar chart (upper part of the figure) shows the number of test cases falling into these categories for each
tool and for each test suite.
To distinguish between various kinds of failures and errors,
we further divided them into different categories (bottom part).

Failures (red) occur either because a correct program is rejected by the type checker (false positive) or
because an ill-typed program is accepted (false negative).
Runtime errors (yellow) include out-of-memory errors, timeouts, or crashes of the type checker.

Rejecting a correct program is either a bug in the tool or a limitation of the type system itself to
type check correct code.
The number of these cases should be kept as small as possible because it prevents type-checking programs that run without errors.
For Dialyzer, this number should be zero as Dialyzer's success typing approach promises to flag only programs that actually fail at runtime.

Accepting an ill-typed program is either a bug in the type checker or a correctness problem of the type system.
Running such a test case leads to a crash that can be traced back to the undetected type error.\footnote{
Another possibility is that a test case has been tailored to document the behavior of the specific tool under test without leading to runtime crashes.
A different tool may successfully type check such a test case.
We manually verified that this situation does not arise; that is, all test cases in the \enquote{false negative}
category definitely lead to runtime crashes.}

Ideally, each tool passes its own test suite (no failed tests) and has no runtime errors.

Dialyzer is a mature tool that handles all test cases without runtime errors.
It fails two test cases expected to pass (false positive);
one of these test cases has been confirmed as a bug and fixed\footnote{\url{https://github.com/erlang/otp/issues/6333}}.
The big downside is that success typing affects its ability to detect type errors reliably:
Dialyzer detects only 89 type errors out of 158; that is, about 44\% of the type errors remain undetected.

We consider Gradualizer also as mature as it has a very low number of runtime errors and is able to handle all Erlang language constructs.
Gradualizer fails to detect about 12\% of the type errors and rejects about 15\% of valid programs across both test suites.
These problems are known and tracked in a special category of the Gradualizer test suite.
It is unknown whether and/or when these test cases will be fixed, but these issues are actively discussed
by the community\footnote{The problematic test cases originate from issues tagged with either \lstinline|discussion| or \lstinline|bug| at \url{https://github.com/josefs/Gradualizer/issues}}.

Etylizer's prototype implementation does not cover all of Erlang's features.
For example, records, maps, or binaries have not been implemented yet, but we believe that these features do not pose significant problems.
Hence, we are optimistic that all tests in the \enquote{not implemented} category will pass in a future release.
Clearly,
Gradualizer's test suite contains several tests for these features, so the number of test cases in the yellow category is high for
this test suite.

For test cases with supported language features, \NAME{} correctly classifies all test cases in its own test suite and about 95\% of the test cases in the Gradualizer test suite.
Testing \NAME{} against the Gradualizer test suite revealed some subtle correctness issues with our tool.
The remaining 5\% of the failed test cases are correctness issues that are tracked in the \NAME{} issue tracker.
Finally, the 14 timeouts and the single out-of-memory error indicate unresolved performance issues.
We have manually reviewed the unexpected results and are confident we can reduce their number to zero with more engineering effort.

\subsection{Performance}

Our experience has shown that even small Erlang definitions cannot be type checked in a reasonable amount of time using a naive implementation of the semantic subtyping approach.
From the start, we needed numerous optimizations to check even trivial functions against their signatures.
While the subtyping implementation is fast, even for function types with many intersection types,
getting the tally algorithm to perform well is challenging.
Tally processes sets of constraint sets, and great care has to be taken to prevent redundant constraint sets from slipping through.
At the same time, the constraint sets have to be kept minimal, otherwise, their size grows exponentially
and type checking becomes practically unfeasible.

Currently, we are able to type check non-trivial functions like \texttt{filtermap} from the Erlang standard library, but this check takes over one minute.
As the implementations of CDuce and similar systems show, it is possible to  speed up the type check further as we have yet to exhaust all possible optimizations, namely hash-consing, lazy evaluation, memoization, and various optimizations of the tally algorithm.
Hints on implementing these optimizations can be found in \cite{journals/corr/abs-1809-01427, Frisch2004}.
%This should cut the time needed by at least one order of magnitude.
Another potential target of optimization
%which are not yet known how to be done is
is the minimization of Boolean algebra terms under subtyping.
This strategy is already solved for Boolean algebra propositional logic \cite{journals/jcss/BuchfuhrerU11}, and the approach and heuristics could prove useful for minimization in our setting.
% In summary, we succeeded in defining a core type system that captures all aspects of the Erlang type specifications.
% Furthermore, the implementation captures all features that were abstracted away by the core languages.
% Thus, the next challenge is getting larger programs to type check quickly.

% optimal data structure for sets of constraint sets which provide efficient implementation of join, meet, and 'has-smaller-constraint' operation

% TODO what to say about type inference?

%%% Local Variables:
%%% mode: latex
%%% TeX-master: t
%%% End:

\section{Related Work}
\label{sec:related-work}

Semantic subtyping has been studied extensively in the programming language community.
Initially, only simple functional languages were defined in terms of set-theoretic types
and there was no support for function types~\cite{conf/popl/pierce01, journals/toit/hosoya03}.
The approach was extended to support type connectives such as
intersection, union, and negation with function, recursive, and product types \cite{conf/icfp/BenzakenCF03, journals/jacm/FrischCB08}.
Next, polymorphism was added~\cite{conf/popl/Castagna0XILP14, conf/popl/Castagna0XA15},
both in an explicitly typed calculus and in a system with local type inference and type
reconstruction.
Finally, typing dynamic parts of programs with gradual typing and refinement of types via occurrence typing
was introduced~\cite{journals/pacmpl/CastagnaL17, journals/pacmpl/CastagnaLNL22}.
While research is ongoing in supporting more features on the type level,  the capabilities already fit
well to the informally specified Erlang type language.
The theory behind set-theoretic typing has, to our knowledge, not yet been applied to
type checking Erlang programs.

The formalism presented in this article is closely related to the work of Castagna and colleagues
on typing polymorphic variants with set-theoretic types~\cite{conf/icfp/CastagnaP016}. Our main extensions of their work are pattern guards
with type tests and support for annotating functions with type signatures. On the practical side, our
implementation extends the set-theoretic type language with Erlang-specific features, which
required changes to the subtyping and tally algorithms.

Over the past decades, various approaches have been tried to verify or statically check Erlang programs.
One common tool is Dialyzer, which is included in every Erlang distribution \cite{conf/erlang/LindahlS05, dialyzer}.
If no type signatures are available, the tool calculates safe over-approximations for the semantics of expressions.
Dialyzer gives the user the guarantee that there no false positives are reported.
If there are type signatures for functions, they will be included in the reasoning process.
The static analysis works reasonably fast for large code bases, scales provably well, and works on all Erlang code.
However, there are various levels of strictness when dealing with type signatures: the programmer must chose between over- or under-approximating type signatures.
It is subtly different from a standard static type system and is therefore unsuited for finding all type errors.
% example polymorphic functions, recursion depth?
% subtle difference example? underspec_in and overspec_out options not possible

%In contrast, recursive types are not handled to their full extent currently in neither Dialyzer nor Gradualizer.
%In Dialyzer, recursive types are unfolded only up to a specified depth, and the bug in Line 12 is missed.
%Gradualizer fails likely for a similar reason.
Gradualizer~\cite{gradualizer} is a more recent tool that aims in a similar direction as we do.
It is able to detect certain classes of errors that Dialyzer is missing.
Still, some things need refinement, most notably the handling of intersection types and subtype polymorphism.
Gradualizer seems to be implemented in an ad-hoc fashion and lacks a published theoretical foundation.
Other type systems~\cite{conf/icfp/MarlowW97, etc, conf/erlang/RajendrakumarB21} lack certain
features and support only a limited subset of Erlang.

Our work shares ties with languages that are dynamically typed but have been later
equipped with a static type system.
Flow and Typescript are typed extensions of the JavaScript language~\cite{journals/pacmpl/ChaudhuriVGRL17, conf/popl/RastogiSFBV15}.
Although they are heavily used in practice, they are less ambitious in their goals as they
only support union and intersection types.
Negation types, which our system requires at least at the meta-theoretic level, are missing.
Further, Typescript trades soundness for scalability.

The theory employed by Ceylon~\cite{conf/oopsla/muehlboeck18}
would also be a good starting point for a type system
for Erlang.
The authors provide a framework for languages with union and intersection types
that can be
adapted to many standard programming languages.
The fact that the framework was designed to be used for nominal languages and
for a constantly changing model (whereas the model for Erlang is fixed)
led us to consider semantic subtyping instead.

MLstruct~\cite{journals/pacmpl/ParreauxC22} supports records, equirecursive types, as well as union, intersection,
and negation. Subtyping is not semantic but algebraic. The language has type inference with principal
types. Our type system also comes with a limited form of inference (it does not infer intersection types)
but lacks principal types. Several design decisions of MLstruct do not seem to be a good fit for Erlang, though.
For example, several arithmetic operators are overloaded in Erlang, while MLstruct does not allow overloading to
be encoded as intersection types. Further, MLstruct treats a union
of a record and a function type equivalently to the top type because MLstruct has no functionality
to distinguish the two parts of the union. Erlang, however, provides this functionality with
the \verb|is_function| predicate.

Typed Racket~\cite{conf/popl/Tobin-HochstadtF08} coined the term \emph{occurrence typing},
meaning that the type of a variable may change, depending on the type tests under which
the variable appears. In Typed Racket, programmers may define custom type tests, whereas
in our system, the set of type tests is fixed.

Work related to typing and verifying the message-passing mechanism in Erlang or Elixir has recently been studied in the context of session types.
As of now, session types have been applied by defining a featherweight Erlang calculus that abstracts the main communication abilities of the language \cite{conf/coordination/MostrousV11}.
Tabone and Francalanza~\cite{DBLP:conf/agere/TaboneF21, DBLP:journals/corr/abs-2208-04631} formalized and implemented another variant of two-party session types for Elixir.
Our work on set-theoretic types for Erlang is largely orthogonal to their works and we expect that \NAME{} can be similarly extended with session types.
Harrison \cite{DBLP:conf/erlang/Harrison18} presents a first version of a static type checker for message compatibility in Erlang programs that can detect orphan messages and possibly dead \lstinline{receive} clauses.
We plan to extend our more general calculus and \NAME{} with similar capabilities.

%%% Local Variables:
%%% mode: latex
%%% TeX-master: "typing-erlang-ifl-2022.tex"
%%% End:

\section{Conclusion and Future Work}
\label{sec:future-work-concl}

This article presented \NAME{}, a static type checker for sequential Erlang based on set-theoretic
types and semantic subtyping. With several examples, we demonstrated that set-theoretic
types are an excellent fit for giving meaning to Erlang's type language and
for checking function signatures against implementations. A formalization of the core calculus
\MinErl{} allowed us to study the type system in great detail. Further, we discussed
and evaluated
an implementation to show the practicality of our approach.

Nevertheless, this article is only the beginning, not the end of our quest to devise a static
type system for Erlang. There are several aspects that we consider as interesting future
work.

Our type checker supports most, but not all of sequential Erlang. For example, support for maps
and records is missing. We plan to add the missing features in the near future.
Also, we need to improve the type checker's performance, and we would
like to add support for type inference of local functions and programs consisting of multiple modules.
On the theoretical side, we would like to prove type soundness and decidability of
type checking for \MinErl{}.

Usually, the migration of larger programs to static type checking happens incrementally. Thus, some parts
of the code are already statically checked, whereas others are not.
Gradual typing\cite{conf/ecoop/SiekT07,conf/erlang/SagonasL08,journals/pacmpl/CastagnaL17,journals/pacmpl/CastagnaLPS19}
would be an interesting addition to our system, as it allows the coexistence of statically
and dynamically typed parts of a program.
Gradual typing could also come to the rescue for parts that cannot be typed
statically (e.g., dynamically loaded code).

Currently, the type checker does not support constructs of distributed Erlang, most notably send and receive.
One option would be to rely on gradual typing; constructs of distributed Erlang would then be
subject to dynamic typing. However, a typical error in distributed Erlang code is sending
a message to a process that the process does not understand.
To capture such errors statically, we would like to implement a rather old
idea of Marlow and Wadler to parametrize the type of a process \cite[Sec.\ 8.2]{conf/icfp/MarlowW97}. With this idea, the type
of process identifiers becomes \lstinline{pid(T)}, where \lstinline{T}
is the type of messages the process accepts. Then, sending a message to a process
with such a type is only allowed if the message's type is a subtype of
\lstinline{T}. On the receiver side, we would implement a simple effect system
to track the kind of messages a process may receive.
With this approach, we could type check simple request-response protocols, but
no complex interactions.
Session types~\cite{conf/coordination/MostrousV11} is an interesting alternative that has also been explored in the context of Erlang.

%%% Local Variables:
%%% mode: latex
%%% TeX-master: "typing-erlang-ifl-2022.tex"
%%% End:

%\begin{acks}
%\end{acks}

\bibliographystyle{ACM-Reference-Format}
\bibliography{typing-erlang-2022}


\begin{thebibliography}{50}

%%% the Web address.

\ifx \showCODEN    \undefined \def \showCODEN     #1{\unskip}     \fi
\ifx \showDOI      \undefined \def \showDOI       #1{#1}\fi
\ifx \showISBNx    \undefined \def \showISBNx     #1{\unskip}     \fi
\ifx \showISBNxiii \undefined \def \showISBNxiii  #1{\unskip}     \fi
\ifx \showISSN     \undefined \def \showISSN      #1{\unskip}     \fi
\ifx \showLCCN     \undefined \def \showLCCN      #1{\unskip}     \fi
\ifx \shownote     \undefined \def \shownote      #1{#1}          \fi
\ifx \showarticletitle \undefined \def \showarticletitle #1{#1}   \fi
\ifx \showURL      \undefined \def \showURL       {\relax}        \fi
% invisible to TeX
\providecommand\bibfield[2]{#2}
\providecommand\bibinfo[2]{#2}
\providecommand\natexlab[1]{#1}
\providecommand\showeprint[2][]{arXiv:#2}

\bibitem[AB(2022)]%
        {ErlangReference2022}
\bibfield{author}{\bibinfo{person}{Ericsson AB}.}
  \bibinfo{year}{2022}\natexlab{}.
\newblock \bibinfo{title}{Erlang Reference Manual, User's Guide, Version
  13.0.3}.
\newblock
\newblock
\newblock
\shownote{\url{https://www.erlang.org/doc/reference_manual/users_guide.html}}.


\bibitem[Abadi and Fiore(1996)]%
        {conf/lics/AbadiF96}
\bibfield{author}{\bibinfo{person}{Mart{\'{\i}}n Abadi} {and}
  \bibinfo{person}{Marcelo~P. Fiore}.} \bibinfo{year}{1996}\natexlab{}.
\newblock \showarticletitle{Syntactic Considerations on Recursive Types}. In
  \bibinfo{booktitle}{\emph{Proceedings, 11th Annual {IEEE} Symposium on Logic
  in Computer Science, New Brunswick, New Jersey, USA, July 27-30, 1996}}.
  \bibinfo{publisher}{{IEEE} Computer Society}, \bibinfo{pages}{242--252}.
\newblock
\urldef\tempurl%
\url{https://doi.org/10.1109/LICS.1996.561324}
\showDOI{\tempurl}


\bibitem[AntidoteDB(2015)]%
        {antidote}
AntidoteDB \bibinfo{year}{2015}\natexlab{}.
\newblock
\newblock
\newblock
\shownote{\url{https://www.antidotedb.eu}}.


\bibitem[Armstrong(2007)]%
        {conf/hopl/Armstrong07}
\bibfield{author}{\bibinfo{person}{Joe Armstrong}.}
  \bibinfo{year}{2007}\natexlab{}.
\newblock \showarticletitle{A history of Erlang}. In
  \bibinfo{booktitle}{\emph{Proceedings of the Third {ACM} {SIGPLAN} History of
  Programming Languages Conference (HOPL-III), San Diego, California, USA, 9-10
  June 2007}}, \bibfield{editor}{\bibinfo{person}{Barbara~G. Ryder} {and}
  \bibinfo{person}{Brent Hailpern}} (Eds.). \bibinfo{publisher}{{ACM}},
  \bibinfo{pages}{1--26}.
\newblock
\urldef\tempurl%
\url{https://doi.org/10.1145/1238844.1238850}
\showDOI{\tempurl}


\bibitem[Benzaken et~al\mbox{.}(2003)]%
        {conf/icfp/BenzakenCF03}
\bibfield{author}{\bibinfo{person}{V{\'{e}}ronique Benzaken},
  \bibinfo{person}{Giuseppe Castagna}, {and} \bibinfo{person}{Alain Frisch}.}
  \bibinfo{year}{2003}\natexlab{}.
\newblock \showarticletitle{CDuce: an XML-centric general-purpose language}. In
  \bibinfo{booktitle}{\emph{Proceedings of the Eighth {ACM} {SIGPLAN}
  International Conference on Functional Programming, {ICFP} 2003, Uppsala,
  Sweden, August 25-29, 2003}}, \bibfield{editor}{\bibinfo{person}{Colin
  Runciman} {and} \bibinfo{person}{Olin Shivers}} (Eds.).
  \bibinfo{publisher}{{ACM}}, \bibinfo{pages}{51--63}.
\newblock
\urldef\tempurl%
\url{https://doi.org/10.1145/944705.944711}
\showDOI{\tempurl}


\bibitem[Bird and Meertens(1998)]%
        {conf/mpc/BirdM98}
\bibfield{author}{\bibinfo{person}{Richard~S. Bird} {and}
  \bibinfo{person}{Lambert G. L.~T. Meertens}.}
  \bibinfo{year}{1998}\natexlab{}.
\newblock \showarticletitle{Nested Datatypes}. In
  \bibinfo{booktitle}{\emph{Mathematics of Program Construction, MPC'98,
  Marstrand, Sweden, June 15-17, 1998, Proceedings}}
  \emph{(\bibinfo{series}{Lecture Notes in Computer Science},
  Vol.~\bibinfo{volume}{1422})}, \bibfield{editor}{\bibinfo{person}{Johan
  Jeuring}} (Ed.). \bibinfo{publisher}{Springer}, \bibinfo{pages}{52--67}.
\newblock
\urldef\tempurl%
\url{https://doi.org/10.1007/BFb0054285}
\showDOI{\tempurl}


\bibitem[Buchfuhrer and Umans(2011)]%
        {journals/jcss/BuchfuhrerU11}
\bibfield{author}{\bibinfo{person}{David Buchfuhrer} {and}
  \bibinfo{person}{Christopher Umans}.} \bibinfo{year}{2011}\natexlab{}.
\newblock \showarticletitle{The complexity of Boolean formula minimization}.
\newblock \bibinfo{journal}{\emph{J. Comput. Syst. Sci.}} \bibinfo{volume}{77},
  \bibinfo{number}{1} (\bibinfo{year}{2011}), \bibinfo{pages}{142--153}.
\newblock
\urldef\tempurl%
\url{https://doi.org/10.1016/j.jcss.2010.06.011}
\showDOI{\tempurl}


\bibitem[Castagna(2018)]%
        {journals/corr/abs-1809-01427}
\bibfield{author}{\bibinfo{person}{Giuseppe Castagna}.}
  \bibinfo{year}{2018}\natexlab{}.
\newblock \showarticletitle{Covariance and Controvariance: a fresh look at an
  old issue (a primer in advanced type systems for learning functional
  programmers)}.
\newblock \bibinfo{journal}{\emph{CoRR}}  \bibinfo{volume}{abs/1809.01427}
  (\bibinfo{year}{2018}).
\newblock
\showeprint[arXiv]{1809.01427}
\urldef\tempurl%
\url{http://arxiv.org/abs/1809.01427}
\showURL{%
\tempurl}


\bibitem[Castagna(2021)]%
        {Castagna2021}
\bibfield{author}{\bibinfo{person}{Giuseppe Castagna}.}
  \bibinfo{year}{2021}\natexlab{}.
\newblock \bibinfo{title}{Programming with union, intersection, and negation
  types}.
\newblock
\newblock
\urldef\tempurl%
\url{https://arxiv.org/abs/2111.03354}
\showURL{%
\tempurl}


\bibitem[Castagna and Frisch(2005)]%
        {conf/ppdp/CastagnaF05}
\bibfield{author}{\bibinfo{person}{Giuseppe Castagna} {and}
  \bibinfo{person}{Alain Frisch}.} \bibinfo{year}{2005}\natexlab{}.
\newblock \showarticletitle{A gentle introduction to semantic subtyping}. In
  \bibinfo{booktitle}{\emph{Proceedings of the 7th International {ACM}
  {SIGPLAN} Conference on Principles and Practice of Declarative Programming,
  July 11-13 2005, Lisbon, Portugal}}, \bibfield{editor}{\bibinfo{person}{Pedro
  Barahona} {and} \bibinfo{person}{Amy~P. Felty}} (Eds.).
  \bibinfo{publisher}{{ACM}}, \bibinfo{pages}{198--199}.
\newblock
\urldef\tempurl%
\url{https://doi.org/10.1145/1069774.1069793}
\showDOI{\tempurl}


\bibitem[Castagna and Lanvin(2017)]%
        {journals/pacmpl/CastagnaL17}
\bibfield{author}{\bibinfo{person}{Giuseppe Castagna} {and}
  \bibinfo{person}{Victor Lanvin}.} \bibinfo{year}{2017}\natexlab{}.
\newblock \showarticletitle{Gradual typing with union and intersection types}.
\newblock \bibinfo{journal}{\emph{Proc. {ACM} Program. Lang.}}
  \bibinfo{volume}{1}, \bibinfo{number}{{ICFP}} (\bibinfo{year}{2017}),
  \bibinfo{pages}{41:1--41:28}.
\newblock
\urldef\tempurl%
\url{https://doi.org/10.1145/3110285}
\showDOI{\tempurl}


\bibitem[Castagna et~al\mbox{.}(2019)]%
        {journals/pacmpl/CastagnaLPS19}
\bibfield{author}{\bibinfo{person}{Giuseppe Castagna}, \bibinfo{person}{Victor
  Lanvin}, \bibinfo{person}{Tommaso Petrucciani}, {and}
  \bibinfo{person}{Jeremy~G. Siek}.} \bibinfo{year}{2019}\natexlab{}.
\newblock \showarticletitle{Gradual typing: a new perspective}.
\newblock \bibinfo{journal}{\emph{Proc. {ACM} Program. Lang.}}
  \bibinfo{volume}{3}, \bibinfo{number}{{POPL}} (\bibinfo{year}{2019}),
  \bibinfo{pages}{16:1--16:32}.
\newblock
\urldef\tempurl%
\url{https://doi.org/10.1145/3290329}
\showDOI{\tempurl}


\bibitem[Castagna et~al\mbox{.}(2022)]%
        {journals/pacmpl/CastagnaLNL22}
\bibfield{author}{\bibinfo{person}{Giuseppe Castagna},
  \bibinfo{person}{Micka{\"{e}}l Laurent}, \bibinfo{person}{Kim Nguyen}, {and}
  \bibinfo{person}{Matthew Lutze}.} \bibinfo{year}{2022}\natexlab{}.
\newblock \showarticletitle{On type-cases, union elimination, and occurrence
  typing}.
\newblock \bibinfo{journal}{\emph{Proc. {ACM} Program. Lang.}}
  \bibinfo{volume}{6}, \bibinfo{number}{{POPL}} (\bibinfo{year}{2022}),
  \bibinfo{pages}{1--31}.
\newblock
\urldef\tempurl%
\url{https://doi.org/10.1145/3498674}
\showDOI{\tempurl}


\bibitem[Castagna et~al\mbox{.}(2015)]%
        {conf/popl/Castagna0XA15}
\bibfield{author}{\bibinfo{person}{Giuseppe Castagna}, \bibinfo{person}{Kim
  Nguyen}, \bibinfo{person}{Zhiwu Xu}, {and} \bibinfo{person}{Pietro Abate}.}
  \bibinfo{year}{2015}\natexlab{}.
\newblock \showarticletitle{Polymorphic Functions with Set-Theoretic Types:
  Part 2: Local Type Inference and Type Reconstruction}. In
  \bibinfo{booktitle}{\emph{Proceedings of the 42nd Annual {ACM}
  {SIGPLAN-SIGACT} Symposium on Principles of Programming Languages, {POPL}
  2015, Mumbai, India, January 15-17, 2015}},
  \bibfield{editor}{\bibinfo{person}{Sriram~K. Rajamani} {and}
  \bibinfo{person}{David Walker}} (Eds.). \bibinfo{publisher}{{ACM}},
  \bibinfo{pages}{289--302}.
\newblock
\urldef\tempurl%
\url{https://doi.org/10.1145/2676726.2676991}
\showDOI{\tempurl}


\bibitem[Castagna et~al\mbox{.}(2014)]%
        {conf/popl/Castagna0XILP14}
\bibfield{author}{\bibinfo{person}{Giuseppe Castagna}, \bibinfo{person}{Kim
  Nguyen}, \bibinfo{person}{Zhiwu Xu}, \bibinfo{person}{Hyeonseung Im},
  \bibinfo{person}{Sergue{\"{\i}} Lenglet}, {and} \bibinfo{person}{Luca
  Padovani}.} \bibinfo{year}{2014}\natexlab{}.
\newblock \showarticletitle{Polymorphic functions with set-theoretic types:
  part 1: syntax, semantics, and evaluation}. In
  \bibinfo{booktitle}{\emph{Proceedings of {POPL}}}.
  \bibinfo{publisher}{{ACM}}, \bibinfo{pages}{5--18}.
\newblock
\urldef\tempurl%
\url{https://doi.org/10.1145/2535838.2535840}
\showDOI{\tempurl}


\bibitem[Castagna et~al\mbox{.}(2016)]%
        {conf/icfp/CastagnaP016}
\bibfield{author}{\bibinfo{person}{Giuseppe Castagna}, \bibinfo{person}{Tommaso
  Petrucciani}, {and} \bibinfo{person}{Kim Nguyen}.}
  \bibinfo{year}{2016}\natexlab{}.
\newblock \showarticletitle{Set-theoretic types for polymorphic variants}. In
  \bibinfo{booktitle}{\emph{Proceedings of the 21st {ACM} {SIGPLAN}
  International Conference on Functional Programming, {ICFP} 2016, Nara, Japan,
  September 18-22, 2016}}, \bibfield{editor}{\bibinfo{person}{Jacques
  Garrigue}, \bibinfo{person}{Gabriele Keller}, {and} \bibinfo{person}{Eijiro
  Sumii}} (Eds.). \bibinfo{publisher}{{ACM}}, \bibinfo{pages}{378--391}.
\newblock
\urldef\tempurl%
\url{https://doi.org/10.1145/2951913.2951928}
\showDOI{\tempurl}


\bibitem[Castagna and Xu(2011)]%
        {conf/icfp/CastagnaX11}
\bibfield{author}{\bibinfo{person}{Giuseppe Castagna} {and}
  \bibinfo{person}{Zhiwu Xu}.} \bibinfo{year}{2011}\natexlab{}.
\newblock \showarticletitle{Set-theoretic foundation of parametric polymorphism
  and subtyping}. In \bibinfo{booktitle}{\emph{Proceedings of {SIGPLAN}}}.
  \bibinfo{publisher}{{ACM}}, \bibinfo{pages}{94--106}.
\newblock
\urldef\tempurl%
\url{https://doi.org/10.1145/2034773.2034788}
\showDOI{\tempurl}


\bibitem[Chaudhuri et~al\mbox{.}(2017)]%
        {journals/pacmpl/ChaudhuriVGRL17}
\bibfield{author}{\bibinfo{person}{Avik Chaudhuri}, \bibinfo{person}{Panagiotis
  Vekris}, \bibinfo{person}{Sam Goldman}, \bibinfo{person}{Marshall Roch},
  {and} \bibinfo{person}{Gabriel Levi}.} \bibinfo{year}{2017}\natexlab{}.
\newblock \showarticletitle{Fast and precise type checking for JavaScript}.
\newblock \bibinfo{journal}{\emph{Proc. {ACM} Program. Lang.}}
  \bibinfo{volume}{1}, \bibinfo{number}{{OOPSLA}} (\bibinfo{year}{2017}),
  \bibinfo{pages}{48:1--48:30}.
\newblock
\urldef\tempurl%
\url{https://doi.org/10.1145/3133872}
\showDOI{\tempurl}


\bibitem[Crary et~al\mbox{.}(1999)]%
        {conf/pldi/CraryHP99}
\bibfield{author}{\bibinfo{person}{Karl Crary}, \bibinfo{person}{Robert
  Harper}, {and} \bibinfo{person}{Sidd Puri}.} \bibinfo{year}{1999}\natexlab{}.
\newblock \showarticletitle{What is a Recursive Module?}. In
  \bibinfo{booktitle}{\emph{Proceedings of {SIGPLAN}}}.
  \bibinfo{publisher}{{ACM}}, \bibinfo{pages}{50--63}.
\newblock
\urldef\tempurl%
\url{https://doi.org/10.1145/301618.301641}
\showDOI{\tempurl}


\bibitem[Damm(1994)]%
        {conf/tacs/Damm94}
\bibfield{author}{\bibinfo{person}{Flemming~M. Damm}.}
  \bibinfo{year}{1994}\natexlab{}.
\newblock \showarticletitle{Subtyping with Union Types, Intersection Types and
  Recursive Types}. In \bibinfo{booktitle}{\emph{Proceedings of {TACS}}}
  \emph{(\bibinfo{series}{LNCS}, Vol.~\bibinfo{volume}{789})}.
  \bibinfo{publisher}{Springer}, \bibinfo{pages}{687--706}.
\newblock
\urldef\tempurl%
\url{https://doi.org/10.1007/3-540-57887-0\_121}
\showDOI{\tempurl}


\bibitem[eqWAlizer(2022)]%
        {eqWAlizer}
eqWAlizer \bibinfo{year}{2022}\natexlab{}.
\newblock
\newblock
\newblock
\shownote{\url{https://github.com/WhatsApp/eqwalizer}}.


\bibitem[Frisch(2004)]%
        {Frisch2004}
\bibfield{author}{\bibinfo{person}{A. Frisch}.}
  \bibinfo{year}{2004}\natexlab{}.
\newblock \emph{\bibinfo{title}{Théorie, conception et réalisation d’un
  langage de programmation adapté à XML}}.
\newblock \bibinfo{thesistype}{Ph.\,D. Dissertation}.
\newblock


\bibitem[Frisch et~al\mbox{.}(2008)]%
        {journals/jacm/FrischCB08}
\bibfield{author}{\bibinfo{person}{Alain Frisch}, \bibinfo{person}{Giuseppe
  Castagna}, {and} \bibinfo{person}{V{\'{e}}ronique Benzaken}.}
  \bibinfo{year}{2008}\natexlab{}.
\newblock \showarticletitle{Semantic subtyping: Dealing set-theoretically with
  function, union, intersection, and negation types}.
\newblock \bibinfo{journal}{\emph{J. {ACM}}} \bibinfo{volume}{55},
  \bibinfo{number}{4} (\bibinfo{year}{2008}), \bibinfo{pages}{19:1--19:64}.
\newblock
\urldef\tempurl%
\url{https://doi.org/10.1145/1391289.1391293}
\showDOI{\tempurl}


\bibitem[Gradualizer(2022)]%
        {gradualizer}
Gradualizer \bibinfo{year}{2022}\natexlab{}.
\newblock
\newblock
\newblock
\shownote{\url{https://github.com/josefs/Gradualizer}}.


\bibitem[Harrison(2018)]%
        {DBLP:conf/erlang/Harrison18}
\bibfield{author}{\bibinfo{person}{Joseph~R. Harrison}.}
  \bibinfo{year}{2018}\natexlab{}.
\newblock \showarticletitle{Automatic detection of core Erlang message passing
  errors}. In \bibinfo{booktitle}{\emph{Proceedings of the 17th {ACM} {SIGPLAN}
  International Workshop on Erlang, {ICFP} 2018, St. Louis, MO, USA, September
  23-29, 2018}}, \bibfield{editor}{\bibinfo{person}{Natalia Chechina} {and}
  \bibinfo{person}{Adrian Francalanza}} (Eds.). \bibinfo{publisher}{{ACM}},
  \bibinfo{pages}{37--48}.
\newblock
\urldef\tempurl%
\url{https://doi.org/10.1145/3239332.3242765}
\showDOI{\tempurl}


\bibitem[Henglein(1993)]%
        {journals/toplas/Henglein93}
\bibfield{author}{\bibinfo{person}{Fritz Henglein}.}
  \bibinfo{year}{1993}\natexlab{}.
\newblock \showarticletitle{Type Inference with Polymorphic Recursion}.
\newblock \bibinfo{journal}{\emph{{ACM} Trans. Program. Lang. Syst.}}
  \bibinfo{volume}{15}, \bibinfo{number}{2} (\bibinfo{year}{1993}),
  \bibinfo{pages}{253--289}.
\newblock
\urldef\tempurl%
\url{https://doi.org/10.1145/169701.169692}
\showDOI{\tempurl}


\bibitem[Hinze(2000)]%
        {journals/jfp/Hinze00}
\bibfield{author}{\bibinfo{person}{Ralf Hinze}.}
  \bibinfo{year}{2000}\natexlab{}.
\newblock \showarticletitle{Perfect trees and bit-reversal permutations}.
\newblock \bibinfo{journal}{\emph{J. Funct. Program.}} \bibinfo{volume}{10},
  \bibinfo{number}{3} (\bibinfo{year}{2000}), \bibinfo{pages}{305--317}.
\newblock
\urldef\tempurl%
\url{https://doi.org/10.1017/s0956796800003701}
\showDOI{\tempurl}


\bibitem[Hosoya and Pierce(2001)]%
        {conf/popl/pierce01}
\bibfield{author}{\bibinfo{person}{Haruo Hosoya} {and}
  \bibinfo{person}{Benjamin Pierce}.} \bibinfo{year}{2001}\natexlab{}.
\newblock \showarticletitle{Regular expression pattern matching for XML}. In
  \bibinfo{booktitle}{\emph{Proceedings of the 28th ACM SIGPLAN-SIGACT
  symposium on Principles of programming languages}}. \bibinfo{pages}{67--80}.
\newblock


\bibitem[Hosoya and Pierce(2003)]%
        {journals/toit/hosoya03}
\bibfield{author}{\bibinfo{person}{Haruo Hosoya} {and}
  \bibinfo{person}{Benjamin~C Pierce}.} \bibinfo{year}{2003}\natexlab{}.
\newblock \showarticletitle{XDuce: A statically typed XML processing language}.
\newblock \bibinfo{journal}{\emph{ACM Transactions on Internet Technology
  (TOIT)}} \bibinfo{volume}{3}, \bibinfo{number}{2} (\bibinfo{year}{2003}),
  \bibinfo{pages}{117--148}.
\newblock


\bibitem[Lindahl and Sagonas(2005)]%
        {conf/erlang/LindahlS05}
\bibfield{author}{\bibinfo{person}{Tobias Lindahl} {and}
  \bibinfo{person}{Konstantinos Sagonas}.} \bibinfo{year}{2005}\natexlab{}.
\newblock \showarticletitle{TypEr: a type annotator of Erlang code}. In
  \bibinfo{booktitle}{\emph{Proceedings of the 2005 {ACM} {SIGPLAN} Workshop on
  Erlang, Tallinn, Estonia, September 26-28, 2005}},
  \bibfield{editor}{\bibinfo{person}{Konstantinos Sagonas} {and}
  \bibinfo{person}{Joe Armstrong}} (Eds.). \bibinfo{publisher}{{ACM}},
  \bibinfo{pages}{17--25}.
\newblock
\urldef\tempurl%
\url{https://doi.org/10.1145/1088361.1088366}
\showDOI{\tempurl}


\bibitem[Lindahl and Sagonas(2006)]%
        {conf/ppdp/LindahlS06}
\bibfield{author}{\bibinfo{person}{Tobias Lindahl} {and}
  \bibinfo{person}{Konstantinos Sagonas}.} \bibinfo{year}{2006}\natexlab{}.
\newblock \showarticletitle{Practical type inference based on success typings}.
  In \bibinfo{booktitle}{\emph{Proceedings of {SIGPLAN}}}.
  \bibinfo{publisher}{{ACM}}, \bibinfo{pages}{167--178}.
\newblock
\urldef\tempurl%
\url{https://doi.org/10.1145/1140335.1140356}
\showDOI{\tempurl}


\bibitem[Marlow and Wadler(1997)]%
        {conf/icfp/MarlowW97}
\bibfield{author}{\bibinfo{person}{Simon Marlow} {and} \bibinfo{person}{Philip
  Wadler}.} \bibinfo{year}{1997}\natexlab{}.
\newblock \showarticletitle{A Practical Subtyping System For Erlang}. In
  \bibinfo{booktitle}{\emph{Proceedings of the 1997 {ACM} {SIGPLAN}
  International Conference on Functional Programming {(ICFP} '97), Amsterdam,
  The Netherlands, June 9-11, 1997}}, \bibfield{editor}{\bibinfo{person}{Simon
  L.~Peyton Jones}, \bibinfo{person}{Mads Tofte}, {and}
  \bibinfo{person}{A.~Michael Berman}} (Eds.). \bibinfo{publisher}{{ACM}},
  \bibinfo{pages}{136--149}.
\newblock
\urldef\tempurl%
\url{https://doi.org/10.1145/258948.258962}
\showDOI{\tempurl}


\bibitem[Mostrous and Vasconcelos(2011)]%
        {conf/coordination/MostrousV11}
\bibfield{author}{\bibinfo{person}{Dimitris Mostrous} {and}
  \bibinfo{person}{Vasco~Thudichum Vasconcelos}.}
  \bibinfo{year}{2011}\natexlab{}.
\newblock \showarticletitle{Session Typing for a Featherweight Erlang}. In
  \bibinfo{booktitle}{\emph{Coordination Models and Languages - 13th
  International Conference, {COORDINATION} 2011, Reykjavik, Iceland, June 6-9,
  2011. Proceedings}} \emph{(\bibinfo{series}{Lecture Notes in Computer
  Science}, Vol.~\bibinfo{volume}{6721})},
  \bibfield{editor}{\bibinfo{person}{Wolfgang~De Meuter} {and}
  \bibinfo{person}{Gruia{-}Catalin Roman}} (Eds.).
  \bibinfo{publisher}{Springer}, \bibinfo{pages}{95--109}.
\newblock
\urldef\tempurl%
\url{https://doi.org/10.1007/978-3-642-21464-6\_7}
\showDOI{\tempurl}


\bibitem[Muehlboeck and Tate(2018)]%
        {conf/oopsla/muehlboeck18}
\bibfield{author}{\bibinfo{person}{Fabian Muehlboeck} {and}
  \bibinfo{person}{Ross Tate}.} \bibinfo{year}{2018}\natexlab{}.
\newblock \showarticletitle{Empowering union and intersection types with
  integrated subtyping}.
\newblock \bibinfo{journal}{\emph{Proceedings of the ACM on Programming
  Languages}} \bibinfo{volume}{2}, \bibinfo{number}{OOPSLA}
  (\bibinfo{year}{2018}), \bibinfo{pages}{1--29}.
\newblock


\bibitem[Parreaux and Chau(2022)]%
        {journals/pacmpl/ParreauxC22}
\bibfield{author}{\bibinfo{person}{Lionel Parreaux} {and}
  \bibinfo{person}{Chun~Yin Chau}.} \bibinfo{year}{2022}\natexlab{}.
\newblock \showarticletitle{MLstruct: principal type inference in a Boolean
  algebra of structural types}.
\newblock \bibinfo{journal}{\emph{Proc. {ACM} Program. Lang.}}
  \bibinfo{volume}{6}, \bibinfo{number}{{OOPSLA2}} (\bibinfo{year}{2022}).
\newblock
\urldef\tempurl%
\url{https://doi.org/10.1145/3563304}
\showDOI{\tempurl}


\bibitem[Pierce(1992)]%
        {Pierce1992}
\bibfield{author}{\bibinfo{person}{Benjamin Pierce}.}
  \bibinfo{year}{1992}\natexlab{}.
\newblock \emph{\bibinfo{title}{Programming with Intersection Types and Bounded
  Polymorphism}}.
\newblock \bibinfo{thesistype}{Ph.\,D. Dissertation}. \bibinfo{address}{USA}.
\newblock


\bibitem[Pierce(2002)]%
        {TAPL}
\bibfield{author}{\bibinfo{person}{Benjamin Pierce}.}
  \bibinfo{year}{2002}\natexlab{}.
\newblock \bibinfo{booktitle}{\emph{Types and Programming Languages}
  (\bibinfo{edition}{1st} ed.)}.
\newblock \bibinfo{publisher}{The MIT Press}.
\newblock
\showISBNx{0262162091}


\bibitem[Pottier and Rémy(2004)]%
        {ATTAPL/typeinference}
\bibfield{author}{\bibinfo{person}{François Pottier} {and}
  \bibinfo{person}{Didier Rémy}.} \bibinfo{year}{2004}\natexlab{}.
\newblock \showarticletitle{The Essence of {ML} Type Inference}.
\newblock In \bibinfo{booktitle}{\emph{Advanced Topics in Types and Programming
  Languages}}, \bibfield{editor}{\bibinfo{person}{Benjamin Pierce}} (Ed.).
  \bibinfo{publisher}{The MIT Press}.
\newblock
\showISBNx{0262162288}


\bibitem[Rajendrakumar and Bieniusa(2021)]%
        {conf/erlang/RajendrakumarB21}
\bibfield{author}{\bibinfo{person}{Nithin~Vadukkumchery Rajendrakumar} {and}
  \bibinfo{person}{Annette Bieniusa}.} \bibinfo{year}{2021}\natexlab{}.
\newblock \showarticletitle{Bidirectional typing for Erlang}. In
  \bibinfo{booktitle}{\emph{Proceedings of the 20th {ACM} {SIGPLAN}
  International Workshop on Erlang, Erlang@ICFP 2021, Virtual Event, Korea,
  August 26, 2021}}, \bibfield{editor}{\bibinfo{person}{Stavros Aronis} {and}
  \bibinfo{person}{Annette Bieniusa}} (Eds.). \bibinfo{publisher}{{ACM}},
  \bibinfo{pages}{54--63}.
\newblock
\urldef\tempurl%
\url{https://doi.org/10.1145/3471871.3472966}
\showDOI{\tempurl}


\bibitem[Rastogi et~al\mbox{.}(2015)]%
        {conf/popl/RastogiSFBV15}
\bibfield{author}{\bibinfo{person}{Aseem Rastogi}, \bibinfo{person}{Nikhil
  Swamy}, \bibinfo{person}{C{\'{e}}dric Fournet}, \bibinfo{person}{Gavin~M.
  Bierman}, {and} \bibinfo{person}{Panagiotis Vekris}.}
  \bibinfo{year}{2015}\natexlab{}.
\newblock \showarticletitle{Safe {\&} Efficient Gradual Typing for TypeScript}.
  In \bibinfo{booktitle}{\emph{Proceedings of the 42nd Annual {ACM}
  {SIGPLAN-SIGACT} Symposium on Principles of Programming Languages, {POPL}
  2015, Mumbai, India, January 15-17, 2015}},
  \bibfield{editor}{\bibinfo{person}{Sriram~K. Rajamani} {and}
  \bibinfo{person}{David Walker}} (Eds.). \bibinfo{publisher}{{ACM}},
  \bibinfo{pages}{167--180}.
\newblock
\urldef\tempurl%
\url{https://doi.org/10.1145/2676726.2676971}
\showDOI{\tempurl}


\bibitem[Riak Core Lite(2020)]%
        {riakcorelite}
Riak Core Lite \bibinfo{year}{2020}\natexlab{}.
\newblock
\newblock
\newblock
\shownote{\url{https://riak-core-lite.github.io}}.


\bibitem[Sagonas(2005)]%
        {dialyzer}
\bibfield{author}{\bibinfo{person}{Konstantinos Sagonas}.}
  \bibinfo{year}{2005}\natexlab{}.
\newblock \showarticletitle{Experience from developing the Dialyzer: A static
  analysis tool detecting defects in Erlang applications}. In
  \bibinfo{booktitle}{\emph{ACM Workshop on the Evaluation of Software Defect
  Detection Tools}}.
\newblock


\bibitem[Sagonas and Luna(2008)]%
        {conf/erlang/SagonasL08}
\bibfield{author}{\bibinfo{person}{Konstantinos Sagonas} {and}
  \bibinfo{person}{Daniel Luna}.} \bibinfo{year}{2008}\natexlab{}.
\newblock \showarticletitle{Gradual typing of erlang programs: a wrangler
  experience}. In \bibinfo{booktitle}{\emph{Proceedings of {SIGPLAN}}}.
  \bibinfo{publisher}{{ACM}}, \bibinfo{pages}{73--82}.
\newblock
\urldef\tempurl%
\url{https://doi.org/10.1145/1411273.1411284}
\showDOI{\tempurl}


\bibitem[Siek and Taha(2007)]%
        {conf/ecoop/SiekT07}
\bibfield{author}{\bibinfo{person}{Jeremy~G. Siek} {and} \bibinfo{person}{Walid
  Taha}.} \bibinfo{year}{2007}\natexlab{}.
\newblock \showarticletitle{Gradual Typing for Objects}. In
  \bibinfo{booktitle}{\emph{Proceedings of {ECOOP}}}
  \emph{(\bibinfo{series}{LNCS}, Vol.~\bibinfo{volume}{4609})}.
  \bibinfo{publisher}{Springer}, \bibinfo{pages}{2--27}.
\newblock
\urldef\tempurl%
\url{https://doi.org/10.1007/978-3-540-73589-2\_2}
\showDOI{\tempurl}


\bibitem[Tabone and Francalanza(2021)]%
        {DBLP:conf/agere/TaboneF21}
\bibfield{author}{\bibinfo{person}{Gerard Tabone} {and} \bibinfo{person}{Adrian
  Francalanza}.} \bibinfo{year}{2021}\natexlab{}.
\newblock \showarticletitle{Session types in Elixir}. In
  \bibinfo{booktitle}{\emph{{AGERE} 2021: Proceedings of the 11th {ACM}
  {SIGPLAN} International Workshop on Programming Based on Actors, Agents, and
  Decentralized Control, Virtual Event / Chicago, IL, USA, 17 October 2021}},
  \bibfield{editor}{\bibinfo{person}{Elias Castegren},
  \bibinfo{person}{Joeri~De Koster}, {and} \bibinfo{person}{Simon Fowler}}
  (Eds.). \bibinfo{publisher}{{ACM}}, \bibinfo{pages}{12--23}.
\newblock
\urldef\tempurl%
\url{https://doi.org/10.1145/3486601.3486708}
\showDOI{\tempurl}


\bibitem[Tabone and Francalanza(2022)]%
        {DBLP:journals/corr/abs-2208-04631}
\bibfield{author}{\bibinfo{person}{Gerard Tabone} {and} \bibinfo{person}{Adrian
  Francalanza}.} \bibinfo{year}{2022}\natexlab{}.
\newblock \showarticletitle{Session Fidelity for ElixirST: {A} Session-Based
  Type System for Elixir Modules}. In \bibinfo{booktitle}{\emph{Proceedings
  15th Interaction and Concurrency Experience, {ICE} 2022, Lucca, Italy, 17th
  June 2022}} \emph{(\bibinfo{series}{{EPTCS}}, Vol.~\bibinfo{volume}{365})},
  \bibfield{editor}{\bibinfo{person}{Cl{\'{e}}ment Aubert},
  \bibinfo{person}{Cinzia~Di Giusto}, \bibinfo{person}{Larisa Safina}, {and}
  \bibinfo{person}{Alceste Scalas}} (Eds.). \bibinfo{pages}{17--36}.
\newblock
\urldef\tempurl%
\url{https://doi.org/10.4204/EPTCS.365.2}
\showDOI{\tempurl}


\bibitem[Tobin{-}Hochstadt and Felleisen(2008)]%
        {conf/popl/Tobin-HochstadtF08}
\bibfield{author}{\bibinfo{person}{Sam Tobin{-}Hochstadt} {and}
  \bibinfo{person}{Matthias Felleisen}.} \bibinfo{year}{2008}\natexlab{}.
\newblock \showarticletitle{The design and implementation of typed scheme}. In
  \bibinfo{booktitle}{\emph{Proceedings of {POPL}}}.
  \bibinfo{publisher}{{ACM}}, \bibinfo{pages}{395--406}.
\newblock
\urldef\tempurl%
\url{https://doi.org/10.1145/1328438.1328486}
\showDOI{\tempurl}


\bibitem[Valliappan(2018)]%
        {etc}
\bibfield{author}{\bibinfo{person}{Nachiappan Valliappan}.}
  \bibinfo{year}{2018}\natexlab{}.
\newblock \emph{\bibinfo{title}{Typing the Untypeable in Erlang-A static type
  system for Erlang using Partial Evaluation}}.
\newblock \bibinfo{thesistype}{Master's\ thesis}.
\newblock


\bibitem[Valliappan and Hughes(2018)]%
        {conf/erlang/ValliappanH18}
\bibfield{author}{\bibinfo{person}{Nachiappan Valliappan} {and}
  \bibinfo{person}{John Hughes}.} \bibinfo{year}{2018}\natexlab{}.
\newblock \showarticletitle{Typing the wild in Erlang}. In
  \bibinfo{booktitle}{\emph{Proceedings of {SIGPLAN}}}.
  \bibinfo{publisher}{{ACM}}, \bibinfo{pages}{49--60}.
\newblock
\urldef\tempurl%
\url{https://doi.org/10.1145/3239332.3242766}
\showDOI{\tempurl}


\bibitem[van Bakel(1995)]%
        {journals/tcs/Bakel95}
\bibfield{author}{\bibinfo{person}{Steffen van Bakel}.}
  \bibinfo{year}{1995}\natexlab{}.
\newblock \showarticletitle{Intersection Type Assignment Systems}.
\newblock \bibinfo{journal}{\emph{Theor. Comput. Sci.}} \bibinfo{volume}{151},
  \bibinfo{number}{2} (\bibinfo{year}{1995}), \bibinfo{pages}{385--435}.
\newblock
\urldef\tempurl%
\url{https://doi.org/10.1016/0304-3975(95)00073-6}
\showDOI{\tempurl}


\end{thebibliography}

\clearpage
\appendix
\section{Algorithmic Typing Rules}
\label{sec:algor-typing-rules}

\Cref{f:constr-gen} and \Cref{f:constr-rew} define rules for generating and simplifying constraints,
also see~\cite[\S\,5]{conf/icfp/CastagnaP016}.
The judgment $\ConstrGen{e}{\Monoty}{\ConstrSet}$ generates a finite set of
constraints $\ConstrSet$ for expression $e$ having type $\Monoty$.
Constraint simplification $\ConstrRew{\Venv}{\ConstrSet}{\SiConstrSet}$ turns
the constraints in $\ConstrSet$ into a finite set
$\SiConstrSet$ with
subtype constraints of the form
$\SubtyConstr{\Monoty}{\Monoty'}$.
The auxiliary relation $\PatTyEnvConstr{\Monoty}{p}{\ConstrSet}{\Venv}$
computes the type environment $\Venv$ for the variables bound by pattern $p$
when used to scrutinize a value of type $\Monoty$. If new type variables are
introduced in $\Venv$, then $\ConstrSet$ constrains them.
The relation $\DefConstrGen{\DefSym}{\ConstrSet}{\Venv}$ generates constraints
$\ConstrSet$ for type checking the body of $\DefSym$; further, it makes the type
to be used for $\DefSym$ available in $\Venv$.

\boxfigSingle{f:constr-gen}{Constraint generation}{
  \[
    \begin{array}{r@{~~}l@{~}r@{~~}l}
      \textrm{constraints} &
      \Constr    & ::= & \SubtyConstr{\Monoty}{\Monoty} \mid
                         \SubtyConstr{x}{\Monoty} \mid
                         \DefConstr{\Venv}{\ConstrSet} \\
      && \mid & \CaseConstr{C}{\Multi{(\InConstr{\Venv_i}{C_i}{\Monoty_i})}} \\
      && \mid & \LetConstr{C}{\Venv}{C}
      \\
      \textrm{finite sets of constraints} & \ConstrSet
    \end{array}
  \]
  \RuleSection{
    \RuleForm{\ConstrGen{e}{\Monoty}{\ConstrSet}}
  }{
    Constraint generation for expressions
  }
  \begin{mathpar}
    \inferrule[c-var]{}{
      \ConstrGen{x}{\Monoty}{\{ \SubtyConstr{x}{\Monoty} \}}
    }

    \inferrule[c-const]{}{
      \ConstrGen{\Const}{\Monoty}{\{ \SubtyConstr{\TyOfConst{\Const}}{\Monoty} \}}
    }

    \inferrule[c-abs]{
      \ConstrGen{e}{\TyvarAux}{\ConstrSet}\\
      \Tyvar, \TyvarAux ~\textrm{fresh}
    }{
      \ConstrGen{\Abs{x}{e}}{\Monoty}{ \{ \DefConstr{\{x : \Tyvar\}}{\ConstrSet},
                                       \SubtyConstr{\Tyvar \to \TyvarAux}{\Monoty} \}}
    }

    \inferrule[c-app]{
      \ConstrGen{e_1}{\Tyvar \to \TyvarAux}{\ConstrSet_1}\\
      \ConstrGen{e_2}{\Tyvar}{\ConstrSet_2}\\
      \Tyvar, \TyvarAux ~\textrm{fresh}
    }{
      \ConstrGen{\App{e_1}{e_2}}{\Monoty}{\ConstrSet_1 \cup \ConstrSet_2 \cup \{\SubtyConstr{\TyvarAux}{\Monoty}\}}
    }

    \inferrule[c-pair]{
      \ConstrGen{e_1}{\Tyvar_1}{\ConstrSet_1}\\
      \ConstrGen{e_2}{\Tyvar_2}{\ConstrSet_2}\\
      \Tyvar_1, \Tyvar_2 ~\textrm{fresh}
    }{
      \ConstrGen{\Pair{e_1}{e_2}}{\Monoty}{
        \ConstrSet_1 \cup \ConstrSet_2 \cup \{\SubtyConstr{\Pairty{\Tyvar_1}{\Tyvar_2}}{\Monoty}\}
      }
    }

    \inferrule[c-case]{
      \Tyvar,\TyvarAux ~\textrm{fresh}\\
      \ConstrGen{e}{\Tyvar}{\ConstrSet}\\
      \ConstrSet' = \ConstrSet \cup \medcup_{i \in I}(\ConstrSet_i \cup \ConstrSet'_i) \cup
                    \{ \SubtyConstr{\Tyvar}{\UnionBig_{i \in I}{\PgUpperTy{\Pg_i}{e}}} \} \\
      \Forall{i \in I}\quad
      \Monoty_i = (\Tyvar \WithoutTy \UnionBig_{j < i}{\PgLowerTy{\Pg_j}{e}}) \Inter \PgUpperTy{\Pg_i}{e}\\
      \PatTyEnvConstr{\Monoty_i}{\PatFromExp{e}}{\ConstrSet_i}{\Venv_i}\\
      \PatTyEnvConstr{\Monoty_i}{\Pg_i}{\ConstrSet'_i}{\Venv'_i}\\
      \Venv_i'' = \Venv_i \InterEnv \Venv'_i\\
      \ConstrGen{e_i}{\TyvarAux}{\ConstrSet''_i}
    }{
      {
        \begin{array}{@{}l@{}}
          \ConstrGen{
            \Case{e}{\Multi{(p_i \to e_i)}}
          }{
            \Monoty\\
          }{
            \{ \CaseConstr{C'}{\Multi{(\InConstr{\Venv''_i}{C''_i}{t_i})}}, \SubtyConstr{\TyvarAux}{\Monoty} \}
          }
        \end{array}
      }
    }

    \infer[c-letrec]{
      \Forall{i \in I}~~ \DefConstrGen{\DefSym_i}{\ConstrSet_i}{\Venv_i}\\
      \Venv = \Venv_{i \in  I}\\
      \ConstrSet = \medcup_{i \in I}\ConstrSet_i\\
      \ConstrGen{e}{\Monoty}{\ConstrSet'}
    }{
      \ConstrGen{
        \Letrec{\Multi{\DefSym}}{e}
      }{
        \Monoty
      }{
        \{ \LetConstr{\ConstrSet}{\Venv}{\ConstrSet'} \}
      }
    }
  \end{mathpar}

  \RuleSection{
    \RuleForm{\DefConstrGen{\DefSym}{\ConstrSet}{\Venv}}
  }{
    Constraint generation for definitions
  }
  \begin{mathpar}
    \inferrule[c-def-annot]{
      \FreeTyVars{\Polyty} = \emptyset\\
      \Polyty = \TyScm{\TyvarSet}{\InterBig_{i \in I}{(\Monoty_i' \to \Monoty_i)}}\\
      \Forall{i \in I}\quad
      \ConstrGen{e}{\Monoty_i}{\ConstrSet_i}
    }{
      \DefConstrGen{\Def{x}{\Polyty}{\Abs{y}{e}}}{
        \{ \DefConstr{ \{y : \Monoty_i'\} }{\ConstrSet_i} \mid i \in I \}
      }{
        \{x : \Polyty \}
      }
    }

    \inferrule[c-def-no-annot]{
      \Tyvar~\textrm{fresh}\\
      \ConstrGen{\Abs{y}{e}}{\Tyvar}{\ConstrSet}
    }{
      \DefConstrGen{\DefNt{x}{\Abs{y}{e}}}{\ConstrSet}{ \{ x : \Tyvar \}}
    }
  \end{mathpar}

  \RuleSection{
    \RuleForm{
      \PatTyEnvConstr{\Monoty}{\Pg}{\ConstrSet}{\Venv}\qquad
      \PatTyEnvConstr{\Monoty}{p}{\ConstrSet}{\Venv}
    }
  }{
    Environments for patterns
  }
  \begin{mathpar}
    \inferrule[c-pg-env]{
      \PatTyEnvConstr{\Monoty}{p}{\ConstrSet}{\Venv}
    }{
      \PatTyEnvConstr{\Monoty}{(\WithGuard{p}{g})}{\ConstrSet}{\Venv \InterEnv \Env{g}}
    }

    \inferrule[c-const-env]{}{
      \PatTyEnvConstr{\Monoty}{\Const}{\emptyset}{\EmptyEnv}
    }

    \inferrule[c-wild-env]{}{
      \PatTyEnvConstr{\Monoty}{\Wildcard}{\emptyset}{\EmptyEnv}
    }

    \inferrule[c-var-env]{}{
      \PatTyEnvConstr{\Monoty}{x}{\emptyset}{\{x : \Monoty \}}
    }

    \inferrule[c-pair-env]{
      \Tyvar_1,\Tyvar_2 ~\textrm{fresh}\\
      \PatTyEnvConstr{\Tyvar_1}{p_1}{\Venv_1}{\ConstrSet_1} \\
      \PatTyEnvConstr{\Tyvar_2}{p_2}{\Venv_2}{\ConstrSet_2}
    }{
      \PatTyEnvConstr{\Monoty}{\Pair{p_1}{p_2}}{
        \ConstrSet_1 \cup \ConstrSet_2 \cup \{ \SubtyConstr{\Monoty}{(\Pairty{\Tyvar_1}{\Tyvar_2})} \}
      }{
        \Venv_1 \InterEnv \Venv_2
      }
    }
  \end{mathpar}
}

\boxfigSingle{f:constr-rew}{Constraint rewriting}{
  \[
    \begin{array}{r@{~}l@{~}r@{~}l@{~~~}l}
      \textrm{simple constraints} &
      \SiConstr    & ::= & \SubtyConstr{\Monoty}{\Monoty} &
      \textrm{finite sets of simple constraints}~ \SiConstrSet
    \end{array}
  \]
  \RuleSection{
    \RuleForm{
      \ConstrRew{\Venv}{\ConstrSet}{\SiConstrSet}  \qquad
      \ConstrRew{\Venv}{\Constr}{\SiConstrSet}
    }
  }{
    Constraint rewriting
  }
  \begin{mathpar}
    \inferrule[rc-set]{
      \Forall{i \in I}~\ConstrRew{\Venv}{\Constr_i}{\SiConstrSet_i}
    }{
      \ConstrRew{\Venv}{\{c_i \mid i \in I\}}{\medcup_{i \in I}{\SiConstrSet_i}}
    }

    \inferrule[rc-subty]{}{
      \ConstrRew{\Venv}{\SubtyConstr{\Monoty}{\Monoty'}}{\{\SubtyConstr{\Monoty}{\Monoty'}\}}
    }

    \inferrule[rc-var]{
      \Venv(x) = \TyScm{\TyvarSet}{\Monoty'}\\
      \textrm{all}~ \Tyvar \in \TyvarSet ~\textrm{fresh}
    }{
      \ConstrRew{\Venv}{\SubtyConstr{x}{\Monoty}}{\{\SubtyConstr{\Monoty'}{\Monoty}\}}
    }

    \inferrule[rc-def]{
      \ConstrRew{\Venv, \Venv'}{\ConstrSet}{\SiConstrSet}
    }{
      \ConstrRew{\Venv}{\DefConstr{\Venv'}{\ConstrSet}}{\SiConstrSet}
    }

    \inferrule[rc-case]{
      \ConstrRew{\Venv}{\ConstrSet}{\SiConstrSet}\\
      \Tysubst \in \Tally[\Free{\Venv}]{\SiConstrSet}\\
      {
        \Forall{i \in I}~
        \begin{cases}
          \SiConstrSet_i = \emptyset &
          \textrm{if}~ \Monoty_i \Tysubst \IsSubty \ErlBot
          \\
          \ConstrRew{\Venv, (\Venv_i\Tysubst)}{\ConstrSet_i}{\SiConstrSet_i} & \textrm{otherwise}
        \end{cases}
      }
    }{
      \ConstrRew{\Venv}{\CaseConstr{\ConstrSet}{\Multi{(\InConstr{\Venv_i}{C_i}{\Monoty_i})}}}{
        \Equiv{\Tysubst} \cup \medcup_{i \in I}\SiConstrSet_i
      }
    }

    \inferrule[rc-let]{
      \ConstrRew{\Venv, \Venv'}{\ConstrSet}{\SiConstrSet}\\
      \Tysubst \in \Tally[\Free{\Venv}]{\SiConstrSet}\\
      \ConstrRew{\Venv, \Gen{\Venv\Tysubst}{\Venv'\Tysubst}}{\ConstrSet'}{\SiConstrSet'}
    }{
      \ConstrRew{\Venv}{\LetConstr{\ConstrSet}{\Venv'}{\ConstrSet'}}{
        \Equiv{\Tysubst} \cup \SiConstrSet'
      }
    }
  \end{mathpar}

  \RuleSection{
    \RuleForm{
      \Gen{\Venv}{\Venv} = \Venv
      \qquad
      \Gen{\Venv}{\Polyty} = \Polyty
      \qquad
      \Equiv{\Tysubst} = \SiConstrSet
    }
  }{
    Auxiliaries
  }

  \begin{mathpar}
    \Gen{\Venv}{ \{ x_i : \Polyty_i \mid i \in I \}} = \{ x_i : \Gen{\Venv}{\Polyty_i} \mid i \in I \}

    \inferrule{
      \TyvarSet \neq \emptyset
    }{
      \Gen{\Venv}{\TyScm{\TyvarSet}{\Monoty}} = \TyScm{\TyvarSet}{\Monoty}
    }

    \inferrule{
      \TyvarSet = \FreeTyVars{\Monoty} \setminus \FreeTyVars{\Venv}
    }{
      \Gen{\Venv}{\Monoty} = \TyScm{\TyvarSet}{\Monoty}
    }

    \Equiv{\Tysubst} = \medcup_{\Tyvar \in \Dom{\Tysubst}} \{
      \SubtyConstr{\Tyvar}{\Tyvar \Tysubst}, \SubtyConstr{\Tyvar \Tysubst}{\Tyvar}
    \}
  \end{mathpar}
}

The solution of a set of subtype constraints $\SiConstrSet$ is a type substitution $\Tysubst$ such that
$\Monoty\Tysubst \IsSubty \Monoty'\Tysubst$ for all
$\SubtyConstr{\Monoty}{\Monoty'} \in \SiConstrSet$.
To compute such a solution, we assume the existence of an algorithm
$\Tally[\TyvarSet]{\SiConstrSet}$
that returns a finite set of type substitutions.
For each $\Tysubst \in \Tally[\TyvarSet]{\SiConstrSet}$ we have
$\Dom{\Tysubst} \cap \TyvarSet = \emptyset$.
The algorithm $\TallySym$ is defined in \cite[\S\,3]{conf/icfp/CastagnaP016}, it is called
\textsf{Sol} there.

In rules \Rule{rc-case} and \Rule{rc-let}, we pick some substitution $\Tysubst$ returned by
$\TallySym$. The choice of $\Tysubst$ is arbitrary, but we must stay consistent with the
choice, ensured by $\Equiv{\Tysubst}$. Further, if the choice of $\Tysubst$ leads to failure,
the algorithm needs to backtrack and picks a different type substitution.

%%% Local Variables:
%%% mode: latex
%%% TeX-master: "typing-erlang-ifl-2022.tex"
%%% End:

\end{document}